\documentclass[twocolumn,prb,notitlepage,floats,superscriptaddress,amsmath,amssymb]{revtex4-2}

\usepackage[version=3]{mhchem} 
\usepackage{bm}
\usepackage[utf8]{inputenc}
\usepackage[T1]{fontenc}
\usepackage{graphicx}
\usepackage{upgreek}
\usepackage{color}
\usepackage{ulem}
\usepackage{hyperref} 
\hypersetup{colorlinks, citecolor=blue, filecolor=blue ,linkcolor=blue , urlcolor=blue, pdftex}
\usepackage{sidecap}
\usepackage{amssymb}
\usepackage{amsmath}
\usepackage[caption=false]{subfig}
\usepackage{multirow}

\def \FUW{University of Warsaw, Faculty of Physics, 02-093 Warsaw, Poland}
\def \Leipzig {Institute for Inorganic Chemistry, Leipzig University, D-04103 Leipzig, Germany}
\def \PW {Institute of Theoretical Physics, Wrocław University of Science and Technology, 50-370 Wrocław, Poland}

\begin{document}

\title{Layered Cr$_2$Ge$_2$Te$_6$ exhibits spin–phonon coupling and isotope-related pseudo-molecular vibrations}

\author{Grzegorz Krasucki}
\email{grzegorz.krasucki@fuw.edu.pl}
\affiliation{\FUW}
\author{Katarzyna Olkowska-Pucko}
\affiliation{\FUW}
\author{Tomasz Wo\'zniak}
\affiliation{\PW}
\author{Mihai~I.~Sturza}
\affiliation{\Leipzig}
\author{Holger Kohlmann}
\affiliation{\Leipzig}
\author{Adam Babi\'nski}
\affiliation{\FUW}
\author{Maciej R. Molas}
\email{maciej.molas@fuw.edu.pl}
\affiliation{\FUW}

\begin{abstract}
The vibrational structure of chromium germanium telluride (Cr$_2$Ge$_2$Te$_6$, CGT) is investigated and a strong spin-phonon coupling is revealed. 
The measured high-resolution Raman scattering (RS) spectra are composed of the 10 Raman-active modes: 5A$_\textrm{g}$ and 5E$_\textrm{g}$, predicted by calculation using the density functional theory and identified using polarization-resolved RS measurements. 
We also studied the effect of temperature on the RS spectra of CGT from 5~K to 300~K.
A strong magneto-phonon coupling in CGT is revealed at temperatures of about 150~K and 60~K, which are associated with the appearance of the local magnetic order in the material and the transition to the complete ferromagnetic phase, respectively.
Moreover, a unique shape of the A$_g^5$ mode composed of a set of very narrow Raman peaks is simulated using a model that takes into account vibrations of Ge-Ge pseudo-molecules for various Ge isotopes.

\end{abstract}

\maketitle

%XXXXXXXXXXXXXXXXXXXXXXXX        INTRO
\section{Introduction \label{sec:Intro}}
Layered magnetic materials (LMMs) have recently been revised for potential applications in spintronics and electronics ~\cite{hoffmann2015, Gibertini2019}. 
Their unique properties of long-range order, combined with their layered structure, uncover new exciting possibilities, both in new magnetic technologies and improving our fundamental understanding of spin dynamics. 
Although many recently studied LMMs, such as FePS$_3$, CrSBr, and MnBi$_2$Te$_4$ are antiferromagnetic, comparatively few members of that family of materials exhibit ferromagnetic interlayer coupling. 
Among the most prominent compounds of ferromagnetic layered are CrCl$_3$, Fe$_3$GeTe$_2$, and Cr$_2$Ge$_2$Te$_6$. 
Their structural stability, compatibility with van der Waals (vdW) heterostructures, and distinctive electronic and magnetic properties make them highly promising candidates for the realization of innovative and advanced device concepts ~\cite{elahi2022}.

Cr$_2$Ge$_2$Te$_6$ (CGT) is a vdW semiconducting ferromagnet with bulk Curie temperature ($T_C$) of about 61 K, a strong magnetic anisotropy with an out-of-plane easy axis, and negligible coercivity~\cite{Carteaux1995, Ji2013, Zhang2016, Gong2017}.
The magnetic properties of CGT can be modified using various perturbations, such as: doping~\cite{Verzhbitskiy2020}, electric field~\cite{Xing2017, Zhuo2021}, strain~\cite{Siskins2022}, and pressure~\cite{Lin2018, Sun2018, Dong2020}.
The CGT is also characterized by a strong magneto-elastic coupling, which can be accompanied by the spin-phonon coupling (SPC).
The common approach to investigating SPC in CGT crystals is the nondestructive and versatile Raman scattering technique~\cite{Tian2016, Samanta2024, Huang2024, Chakkar2024, Idzuchi2025}.

In this work, we reveal a complete set of the 10 theoretically predicted Raman active modes, 5A$_\textrm{g}$ and 5E$_\textrm{g}$, using high-resolution Raman scattering (RS) spectroscopy performed with polarization resolution and as a function of temperature from 5~K to 300 K.
Our results demonstrate that nearly all observable modes exhibit distinct phase-dependent changes in their temperature-dependent properties.
A strong magneto-phonon coupling in CGT is revealed at temperatures of about 150~K and 60~K, which are associated with the appearance of the local magnetic order in the material and the transition to the complete ferromagnetic phase, respectively.
We also successfully extracted spin–phonon coupling constants for almost all detected modes. 
Furthermore, we observe an unusual fine structure of the A$_g^5$ mode comprising a set of very narrow peaks, which is reconstructed using a model that takes into account vibrations of Ge-Ge pseudo-molecules for various Ge isotopes.

%XXXXXXXXXXXXXXXXXXXXXXXX        EXPERIMENTAL RESULTS
\section{Results and discussion \label{sec:result}}
\subsection*{Crystal structure and phonon dispersion \label{sec:crystal}}

\begin{figure*}[t]
		\subfloat{}%
		\centering
        \includegraphics[width=1\linewidth]{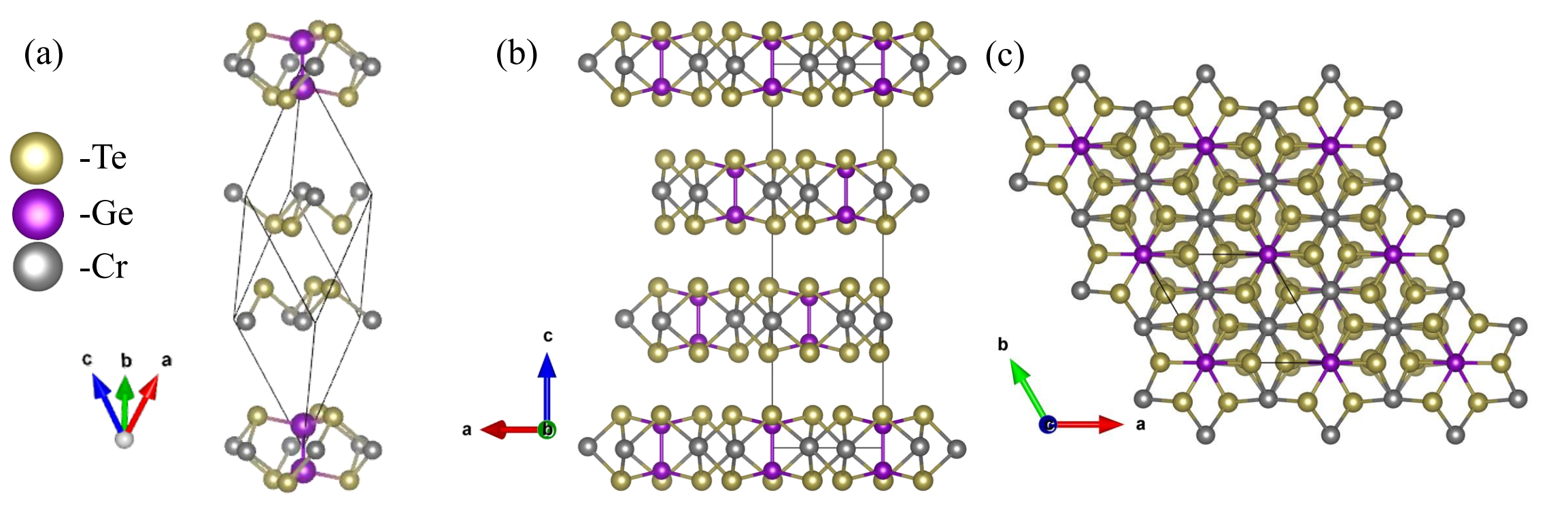}
        \caption {The schematic representation of (a) perspective with rombohedral axis, (b) side, and (c) top views with hexagonal axis of the atomic structure of the CGT crystal. 
        The black shapes represent the unit cells. }
		\label{Crystal_s}
\end{figure*}

%Plik CIF z tego artykułu 
%Carteaux, V.; Brunet, D.; Ouvrard, G.; Andre, G.
%Crystallographic, magnetic and electronic structures of a new layered ferromagnetic compound Cr2 Ge2 Te6
%Journal of Physics: Condensed Matter, 1995, 7, 69-87

CGT crystallizes in a rhombohedral structure (space group R$\bar{3}$, no. 148) with a vdW stacking of Te-Ge-Cr-Ge-Te hexagonal layers~\cite{Carteaux1995,Ji2013}.
The schematic representation of the atomic structure of the CGT crystal is shown in Fig.~\ref{Crystal_s}. 
CGT crystals are composed of a series of octahedral units in which Te atoms are located at the vertices, and either a single Cr atom or two Ge atoms are positioned at the center. 
These atoms are bonded by covalent bonds.
The octahedra connect with each other to form a single layer of material, as shown in Fig.~\ref{Crystal_s}.
Between the individual layers of the crystal, there are only weak vdW forces, which makes this material particularly well-suited for mechanical exfoliation.
The octahedra containing Cr and Ge atoms alternate, forming a repeating supercell.
The entire crystal contains an equal number of Cr and Ge atoms.
However, since the octahedra with Ge contain two Ge atoms each, this is compensated by having twice as many Cr-containing units as Ge-containing ones throughout the crystal.

\begin{figure}[b]
		\subfloat{}%
		\centering
		\includegraphics[width=1 \linewidth]{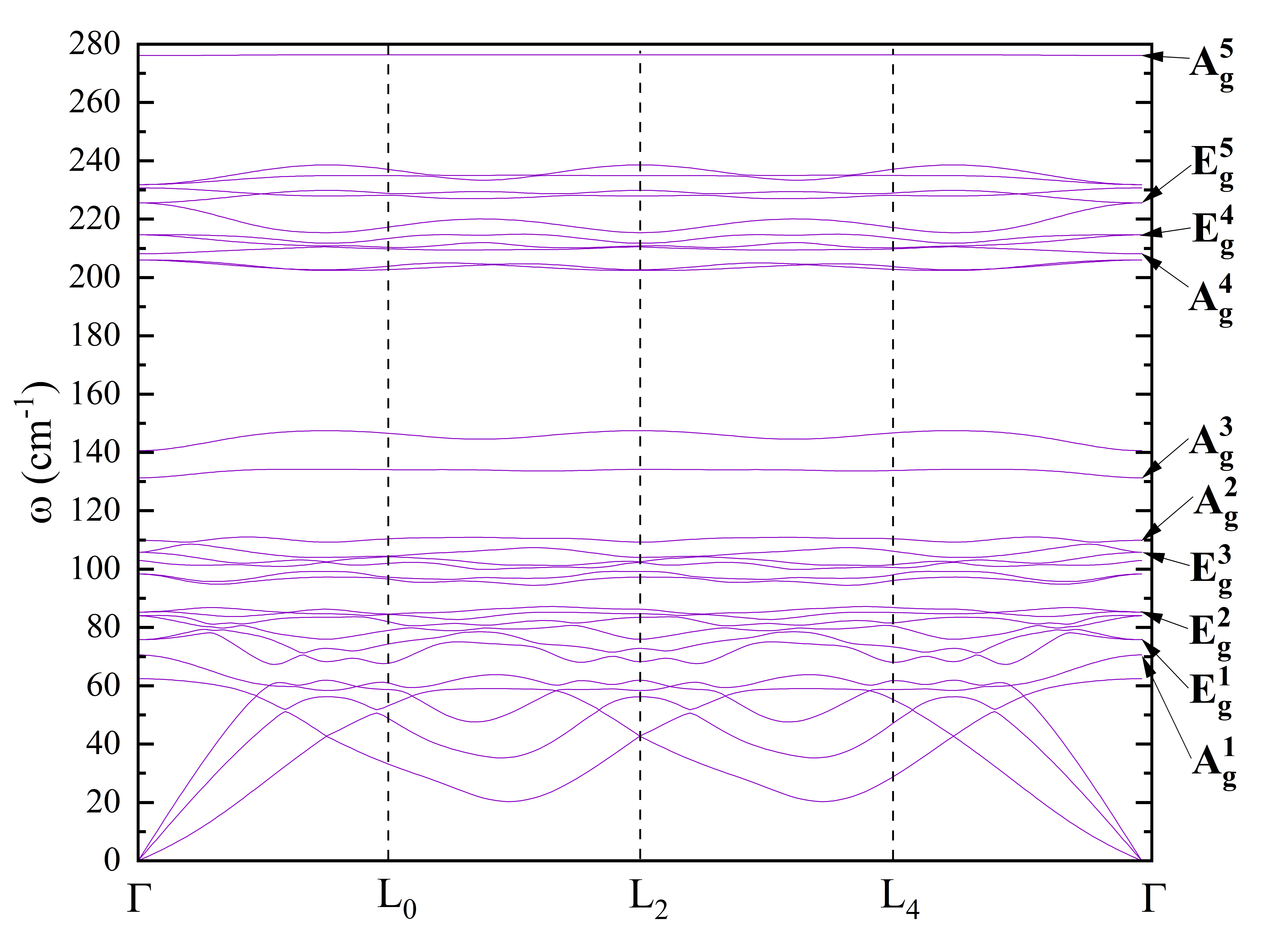}
    	\caption{Phonon dispersion of bulk CGT crystal in rhombohedral primitive cell with ferromagnetic order. 
            The $\Gamma$ point corresponds to center of the Brillouin zone. 
            The blue curves denotes Raman-active modes, while the orange curves correspond to the remaining modes, i.e. infrared-active or inactive.}
		\label{fig:2}
\end{figure}

The crystal structure of the CGT belongs to space group no. 148 with Cr, Ge, and Te atoms in the 6c, 6c, and 18f Wyckoff positions, respectively. 
The symmetries of the lattice vibrations at the $\Gamma$ point of the Brillouin zone (BZ) can be assigned to the following 30 irreducible representations: $\Gamma \equiv 5\textrm{A}_\textrm{g} \oplus 5\textrm{E}_\textrm{g} \oplus 5\textrm{A}_\textrm{u} \oplus 5\textrm{E}_\textrm{u}$.  
There are 15 Raman-active phonon modes: $5\textrm{A}_\textrm{g}$ and $10\textrm{E}_\textrm{g}$, and 15 infrared-active ones: $5\textrm{A}_\textrm{u}$ and $10\textrm{E}_\textrm{u}$.
Note that the $\textrm{E}_\textrm{g}$ and $\textrm{E}_\textrm{u}$ modes are doubly degenerate, which gives rise to 10 Raman- and infrared-active vibrations.
We optimized the geometric parameters and calculated the dispersion of the phonon modes for bulk CGT in the presence of ferromagnetic order within the density functional theory (DFT) framework, as illustrated in Fig.~\ref{fig:2}.
The optimized lattice constant of the rhombohedral cell, $a$=7.889~\AA, corresponds to $a$=6.882~\AA~in the hexagonal cell, in good agreement with the experimental value of $a$=6.823~\AA ~and $c$=20.564~\AA~\cite{Ji2013}.
The Raman-active modes at the $\Gamma$ point are labeled on the right with the superscripts on the labels, which describe additional numbering because of their increased Raman shift. 
In particular, the A$_g^5$ mode shows a negligible dependence on the momentum ($k$) throughout the BZ, compared to the other modes.

\subsection*{Polarization dependence of phonon modes \label{sec:polarization}}

\begin{figure}[t]
		\subfloat{}%
		\centering
		\includegraphics[width=1 \linewidth]{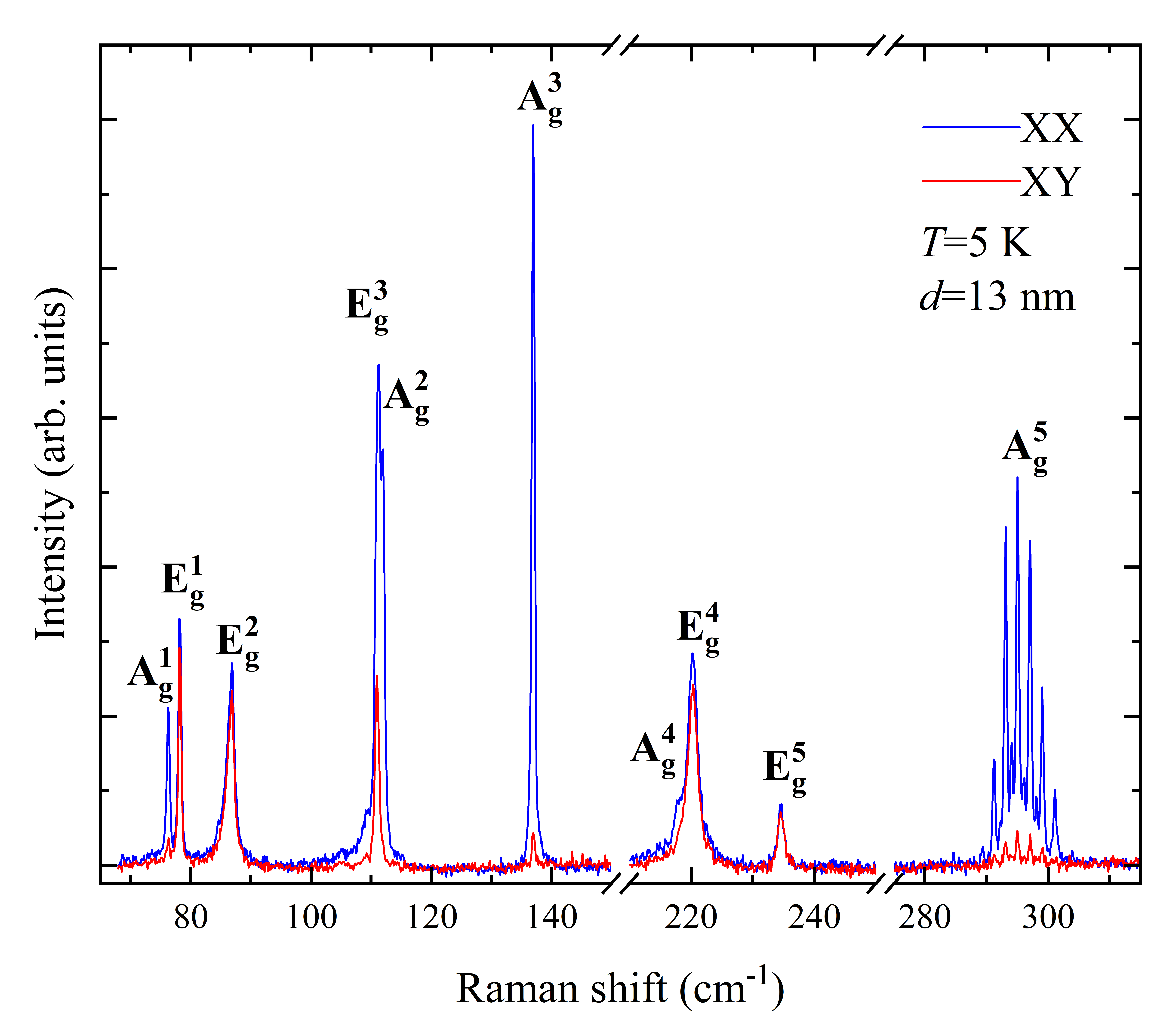}
        \caption{Raman Scattering spectra of the 24 nm thick CGT flake measured at 5~K with co-linear (XX) and cross-linear (XY) polarization using excitation energy 1.58~eV and excitation power of 1~mW.
        The horizontal scale was adjusted with breaks for clarity.}
		\label{fig:1}
\end{figure}

We investigated spin-phonon coupling using the RS technique on mechanically exfoliated thick CGT layers with thicknesses in the range of tens of nanometers deposited on Si/SiO$_2$, which can be treated as bulk-like CGT flakes.
The optical photographs of the flakes under study with their corresponding atomic force microscopy (AFM) images are shown in Section S1 of the Supporting Information (SI).
The representative low temperature ($T$=5~K) polarized RS spectra
of a 24~nm thick CGT flake with co-linear (XX) and cross-linear (XY) polarization using excitation energy 1.58~eV are presented in Fig.~\ref{fig:1}.
The choice of excitation used in the following analysis is motivated by the quality of the Raman spectrum obtained under illumination of 1.58~eV as compared to other available energies, $i.e.$ 1.96~eV, 2.21~eV, 2.41~eV, and 3.06~eV, see Section S2 of the SI for details.
Ten distinct RS peaks are observed in the spectra, as predicted theoretically.
Due to the observed polarization-dependences of peaks in the Figure (see also Section S3 of the SI for details), they can be attributed to the five out-of-plane $\textrm{A}_\textrm{g}$ modes, apparent at 76.5~cm$^{-1}$ ($\textrm{A}_\textrm{g}^1$), 111.6~cm$^{-1}$ ($\textrm{A}_\textrm{g}^2$), 137.1~cm$^{-1}$ ($\textrm{A}_\textrm{g}^3$), 219.0~cm$^{-1}$ ($\textrm{A}_\textrm{g}^4$), and 295.1~cm$^{-1}$ ($\textrm{A}_\textrm{g}^5$); and five in-plane $\textrm{E}_\textrm{g}$ modes, apparent at 78.5~cm$^{-1}$ ($\textrm{E}_\textrm{g}^1$), 86.8~cm$^{-1}$ ($\textrm{E}_\textrm{g}^2$), 111.2~cm$^{-1}$ ($\textrm{E}_\textrm{g}^3$), 220.4~cm$^{-1}$ ($\textrm{E}_\textrm{g}^4$), and 234.6~cm$^{-1}$ ($\textrm{E}_\textrm{g}^5$). 
DFT calculations yield frequencies: for $\textrm{A}_\textrm{g}$ modes, 70.6~cm$^{-1}$, 109.9~cm$^{-1}$, 131.3~cm$^{-1}$, 208.1~cm$^{-1}$, and 276.1~cm$^{-1}$;  for $\textrm{E}_\textrm{g}$ modes, 75.8~cm$^{-1}$, 85.3~cm$^{-1}$, 105.8~cm$^{-1}$, 214.6~cm$^{-1}$, and 225.6~cm$^{-1}$.
The experimental and theoretical phonon energies show good overall agreement. 
Although the calculated values are slightly lower than those extracted from the RS spectra, their relative energy differences align very well.
As the $\textrm{E}^3_\textrm{g}$ and $\textrm{A}^2_\textrm{g}$ modes exhibit nearly degenerate energies at 5~K, we analyzed their polarization evolutions at room temperature to confirm their distinct identities, see Section S3 in the SI for details.
In particular, the $\textrm{A}^5_\textrm{g}$ mode displays an unexpected multipeak structure, which will be discussed in detail in the final section of this work.

\subsection*{Phase transitions revealed by temperature-dependent Raman scattering \label{sec:temperature}}

\begin{figure}[t]
		\subfloat{}%
		\centering
		\includegraphics[width=1 \linewidth]{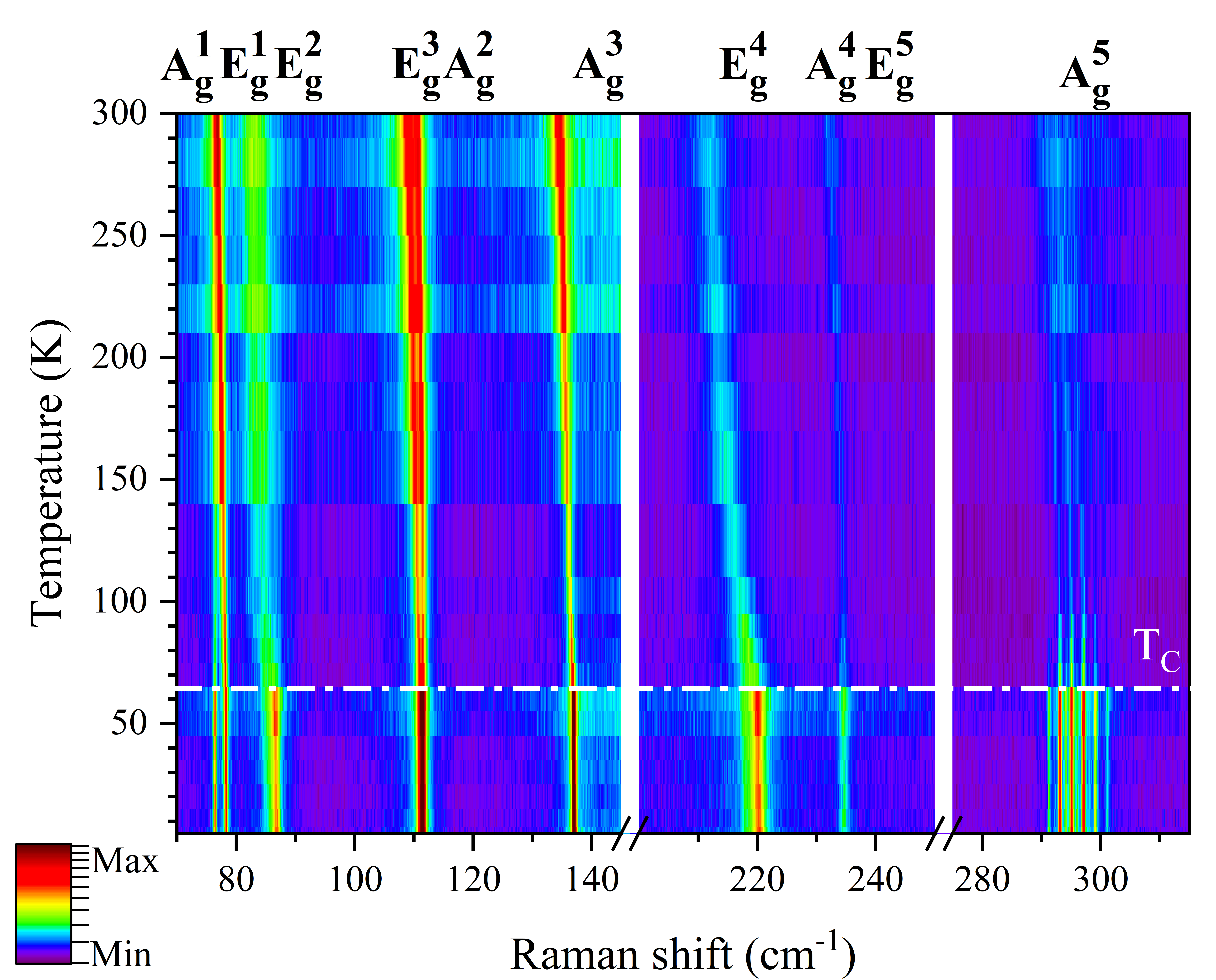}
        \caption{False-color map of the temperature evolution of the RS spectra of the exfoliated 24-nm thick CGT flake under 1.58~eV excitation with 1~mW power.
        The white horizontal dash-dot line denotes Curie temperature ($T_C$) of CGT. 
        The horizontal scale was adjusted with breaks for clarity.}
		\label{fig:3}
\end{figure}

To investigate the coupling between the magnetic and vibrational properties of CGT, we performed temperature-dependent RS measurements.
Fig.~\ref{fig:3} shows the false-color map of the temperature evolution of the RS spectra of the investigated CGT flake.
Interestingly, the RS spectra remain remarkably stable with minimal variations in energies, linewidths, and intensities of the Raman peaks up to temperatures of about 60~K, which correspond to the $T_C$ temperature of CGT.
At higher temperatures ($T>T_C$), a pronounced transformation in the RS spectra is apparent.
A rapid decrease in mode energies and a significant reduction in their intensities are easily recognized. 
The most spectacular effect is observed for the $\textrm{A}_\textrm{g}^1$ peak, whose energy exhibits a significant non-monotonic dependence in the temperature range from $T_C$ to 300~K.
This pronounced spectral reorganization in the RS spectra can be attributed to the significant spin-phonon coupling in CGT due to the transition from the ferromagnetic to the paramagnetic phase when passing through the Curie temperature, consistent with previous reports devoted to this material~\cite{Tian2016, Chakkar2024, Samanta2024}.
Note that because of the much higher spectral resolution in our experiments as well as the high quality of the investigated CGT crystals, our temperature-dependent results are much more detailed than the previously published~\cite{Tian2016, Chakkar2024, Samanta2024}.

\begin{figure*}[!th]
		\subfloat{}%
		\centering
		\includegraphics[width=1 \linewidth]{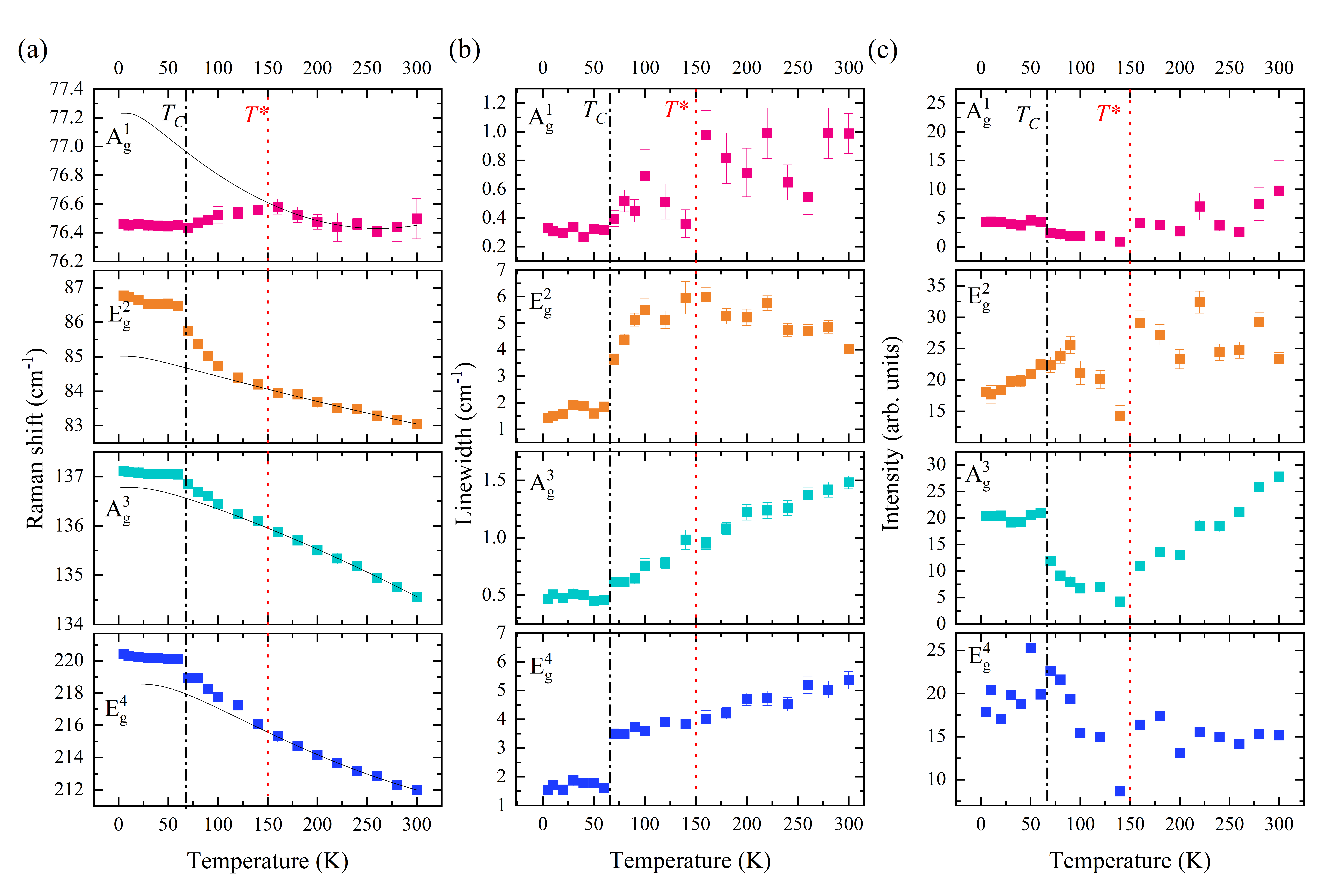}
        \caption{Temperature evolutions of the (a) Raman shifts, (b) linewidths and (c) integrated intensities of the selected four Raman modes, $i.e.$ A$_g^1$, E$_g^2$, A$_g^3$, E$_g^4$ measured on the exfoliated 24-nm thick CGT flake.
        The black vertical dash-dot line corresponds to the Curie temperature ($T_C$), while the red vertical doted line denote the second phase change temperature ($T^*$). 
        The gray curves are result of fitting using Balkanski model (Eg.~\ref{balkanski}) to the data measured only above the $T_1$, $i.e.$ in the paramagnetic phase.}
		\label{fig:4}
\end{figure*}

To complete our temperature-related analysis, we extracted the energies, linewidths, and intensities of all the observed phonon modes by their deconvolution using the Lorentzian function. 
The obtained temperature evolutions of the peak energies, linewidths (full widths at half maximum, FWHMs), and intensities 
for all 10 Raman peaks are presented in Section S4 of the SI.
For clarity, here, we focus on the temperature evolution of the $\textrm{A}_\textrm{g}^1$, $\textrm{E}_\textrm{g}^2$, $\textrm{A}_\textrm{g}^3$, and $\textrm{E}_\textrm{g}^4$ modes, as shown in Fig.~\ref{fig:4}, which displays their energy, linewidth, and intensity dependencies on temperature.

Firstly, let us focus on the energy dependence of the studied Raman peaks as a function of temperature, see Fig.~\ref{fig:4}(a).
Below $T_C$ value, the $\textrm{A}_\textrm{g}^1$, $\textrm{E}_\textrm{g}^2$, $\textrm{A}_\textrm{g}^3$, and $\textrm{E}_\textrm{g}^4$ energies remain nearly constant.
In the intermediate temperature range (60–150~K), their energy evolutions deviate from almost linear shifts common to various layered materials~\cite{Lapinska2016, Buruiana2022, Muhammad2024}.
In particular, the $\textrm{A}_\textrm{g}^1$ peak experiences a blueshift with increased temperature up to about 150~K, when its dependence changes to the redshift. 
This may indicate that the transition from the ferromagnetic phase to the paramagnetic phase is not abrupt when passing through the Curie temperature.
We assume that an intermediate state with local ferromagnetic domains occurs at temperatures between $T_C$$\approx$60~K and $T^*$$\approx$150~K~\cite{Chakkar2024}, while a pure paramagnetic phase is evident at higher temperatures. 

The measured temperature dependence of the phonon frequencies generally contains contributions from both explicit anharmonic phonon–phonon interactions and implicit anharmonicity associated with thermal expansion. Spachmann et al. reported strongly anisotropic thermal expansion in Cr$_2$Ge$_2$Te$_6$, with effective Grüneisen parameters of $\gamma_{ab}$ = 0.32 and $\gamma_c$ = 0.56 ~\cite{Spachmann2022_rev}. Using the reported lattice expansion, we estimate that the corresponding quasi-harmonic contribution produces a fractional phonon-frequency change of the order of several 10$^{-3}$ over the investigated temperature range. We note that these values represent effective macroscopic rather than mode-specific Grüneisen parameters and therefore provide only an order-of-magnitude estimate. Moreover, mechanical coupling of the exfoliated flakes to the substrate may modify their thermal expansion. Accordingly, the Balkanski model used here should be regarded as an effective description of the dominant anharmonic temperature dependence, while a smaller contribution from lattice thermal expansion cannot be completely excluded.

To quantitatively analyze the spin-phonon coupling in CGT, we fitted the temperature evolution of the phonon energies, $\omega_{anh}(T)$, using the anharmonic model proposed in Ref.~\citenum{Balkanski1983}, which reads: 
\begin{equation}
\begin{aligned}
    \omega_{anh}(T) = \omega_0 + A \left(\frac{1}{e^{x}+1} \right) \\ + B \left(1 + \frac{3}{e^{y}-1} + \frac{3}{(e^{y}-1)^2}  \right),
\end{aligned}
    \label{balkanski}
\end{equation}
where $\omega_0$, $A$, and $B$ are fitting parameters, $x = \frac{\hbar\omega_{0}}{2k_{B}T}$, $y = \frac{\hbar\omega_{0}}{3k_{B}T}$, and $\omega_{0}+A+B$ is the phonon frequency at 0~K.

As can be seen in Fig.~\ref{fig:4}(a), this model can characterize the phonon temperature dependences, but only for temperatures higher than $T^*$.
At $T<T_\textrm{C}$ the phonon energies are significantly smaller or larger than those predicted by the model. 
The reason for this aspect is the spin-phonon coupling caused by ionic motions, as in other LMMs~\cite{Kozlenko2021}. 
The spin-phonon coupling coefficient is given by the formula~\cite{Tian2016, Chakkar2024}: 
\begin{equation}
\omega(T) \sim \omega_{anh}(T) + \lambda \langle S_{i} \cdot S_{j} \rangle 
\label{spin}
\end{equation} 
where $\omega(T)$ represents the measured phonon energy, $\omega_{anh}(T)$ denotes the phonon energy without spin-phonon coupling, $\lambda$ denotes the coupling strength, $\langle S_{i} \cdot S_{j}  \rangle$ is the spin-spin correlation function of neighboring spins, and the values of $S_{i}$ and $S_{j}$ are 3/2~\cite{Ji2013, Han2019, Koo2022} giving $\langle S_{i} \cdot S_{j} \rangle$ value of 9/4. 
The obtained $\lambda$ values for the $\omega_{anh}$($T$=5~K) are -0.34~cm$^{-1}$, 0.78~cm$^{-1}$, 0.15~cm$^{-1}$, and 0.82cm~$^{-1}$
correspondingly for the $\textrm{A}_\textrm{g}^1$, $\textrm{E}_\textrm{g}^2$, $\textrm{A}_\textrm{g}^3$, and $\textrm{E}_\textrm{g}^4$ modes. 

The varying magnitude of phonon renormalization observed across different modes near 60~K and 150~K can be understood through mode-selective spin-phonon coupling ~\cite{Singh2019_rev}. Because each phonon mode is characterized by a distinct eigenvector, different modes dynamically modulate the relevant magnetic exchange pathways, bond lengths, and bond angles to different extents. Modes involving stronger modulation of the dominant magnetic exchange pathways are therefore expected to exhibit larger spin-phonon coupling coefficients $\lambda$, and consequently more pronounced frequency renormalization. Furthermore, the anomalies observed around 150~K can be associated with the development of short-range spin correlations in the correlated paramagnetic regime, which may selectively affect phonon modes sensitive to local exchange pathways before the onset of long-range ferromagnetic ordering near 60~K.

Some of our results are consistent with previously reported findings~\cite{Tian2016}. 
We note that the literature contains contradictory results for comparable measurements but performed with much smaller spectral resolution~\cite{Chakkar2024}. Furthermore, for most of the phonon modes analyzed in this work, the spin–phonon coupling constants have not been previously reported. 
The extracted $\lambda$ values for all the investigated modes compared with those reported in Refs.~\citenum{Tian2016, Chakkar2024} are summarized in a table in Section S4 of the SI.

The influence of the spin-phonon coupling is also displayed in the temperature evolution of the Raman peaks, see Fig. \ref{fig:4}(b).
The linewidths of the $\textrm{A}_\textrm{g}^1$, $\textrm{E}_\textrm{g}^2$, $\textrm{A}_\textrm{g}^3$, and $\textrm{E}_\textrm{g}^4$ modes maintain a consistent level up to $T_C$ value. 
All the analyzed linewidths reveal distinct changes upon crossing the Curie temperature, $i.e.$ the modes broaden significantly. 
In particular, the $\textrm{E}_\textrm{g}^2$ and $\textrm{E}_\textrm{g}^4$ linewidths increase more than two times at $T_C$.
The signature of the $T^*$ temperature is seen for the A$_g^1$ mode, while for nine other modes (see Section S4 of the SI), there is no obvious indication.
Note that our results are qualitatively in agreement with the previous reports~\cite{Tian2016, Chakkar2024}, but there is an evident difference in the extracted linewidths of phonon modes.
In our experiment, the $\textrm{A}_\textrm{g}^3$ linewidth is of about 0.5~cm$^{-1}$ at 5~K, while the corresponding values reported in the literature are of around 0.7~cm$^{-1}$~\cite{Tian2016} and 2.6~cm$^{-1}$~\cite{Chakkar2024}.
This indicates that the quality of the investigated CGT crystal in this work is much higher or that the resolution of our experimental setup is substantially larger than that of the existing research.

The temperature evolutions of the phonon intensities are particularly intriguing, as two competing processes may influence it.
There are non-resonant and resonant conditions of the RS excitation, which change as a function of temperature.
Moreover, the temperature-induced transition between the ferromagnetic and paramagnetic phases may also affect the Raman modes intensities.
Consequently, it is very difficult to determine independently both effects, but we describe the apparent signatures of the magnetic one in the following.
The $\textrm{E}_\textrm{g}^4$ intensity exhibits an almost fixed value with a fairly large dispersion, whereas those of the $\textrm{A}_\textrm{g}^1$, $\textrm{E}_\textrm{g}^2$, and $\textrm{A}_\textrm{g}^3$ modes are described by more organized trends.
There are step-like intensity variations at $T_C$ ($\textrm{A}_\textrm{g}^1$ and $\textrm{E}_\textrm{g}^2$) and at $T^*$ ($\textrm{A}_\textrm{g}^1$, $\textrm{E}_\textrm{g}^2$, and $\textrm{A}_\textrm{g}^3$), which are assigned to the transition between the ferromagnetic and (quasi-)paramagnetic phases.
Our results indicate that the intensity can also be used to distinguish the transition temperature between different magnetic phases, as reported for other LMMs, $e.g.$ CrBr$_3$~\cite{Kipczak2024} and CrSBr~\cite{Pawbake2023}.

\begin{figure}[!t]
		\subfloat{}%
		\centering
		\includegraphics[width=1\linewidth]{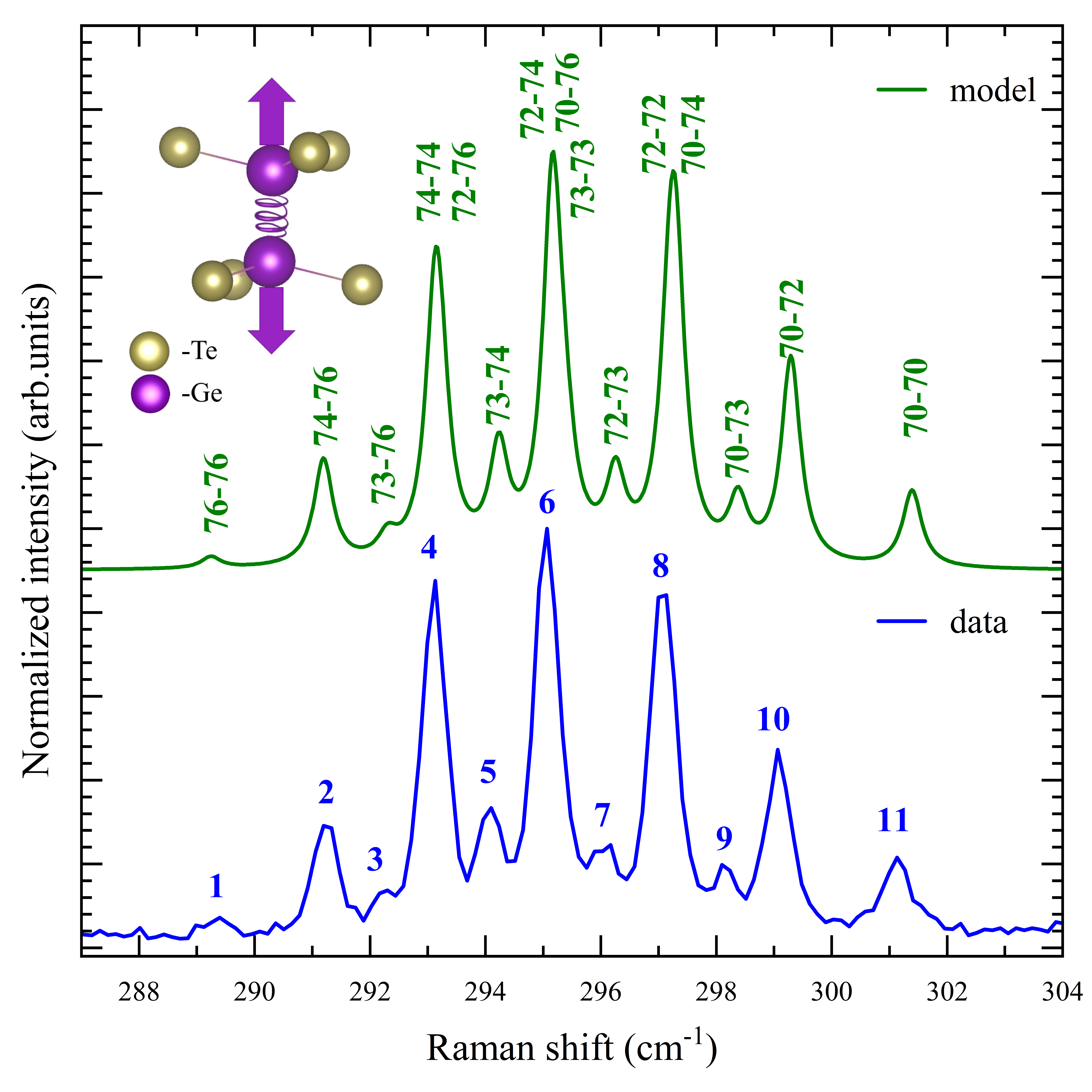}
        \caption{Experimental Raman spectrum of the A$_g^5$ mode measured on the exfoliated 24-nm thick CGT flake at $T$=5~K using excitation energy 1.58~eV and excitation power of 1~mW (blue curve) and simulated Raman spectrum of the A$_g^5$ mode including various Ge isotopes (green curve).
        The 1$\dots$11 numbers label the phonon peaks observed in experimental data, while the corresponding calculated lines are indicated by their assignment to pairs formed by different Ge isotopes ($e.g.$ 73-76).
        The inset shows the schematic representation of the A$_g^5$ vibration.}
		\label{fig:5}
\end{figure}

\subsection*{Fine structure of Ge-Ge vibration \label{sec:fine}}
As it can be seen in Fig.~\ref{fig:5}, the A$_\textrm{g}^5$ peak has a very specific landscape. 
This mode exhibits a structure composed of eleven narrow peaks, which we labeled using numbering because of their increased Raman shift. 
The energy distance difference the consecutive peaks is only about 1~cm$^{-1}$.
% 1.9652$\pm$ 0.0059~cm$^{-1}$.
The linewidths of the individual peaks are on the order of 0.4~cm$^{-1}$ and are comparable to the smallest linewidths for other phonon modes, $i.e.$ A$_\textrm{g}^1$, A$_\textrm{g}^2$, A$_\textrm{g}^3$, E$_\textrm{g}^1$. %0.382$\pm$0.018~cm$^{-1}$
Importantly, the A$_\textrm{g}^5$ fine structure is observable at $T$=5~K under specific excitation energies, 1.58~eV and 1.91~eV, likely due to the higher spectral resolution of the measurement setup in that energy range, see Section S2 of the SI. 
This fine structure is present throughout the ferromagnetic phase (from 5~K to 60~K).
The modes characterized by the lowest intensity, $i.e.$ 1, 3, 5, 7, 9 disappear abruptly above the $T_C$ temperature, while those of the remaining ones can be resolved up to almost the $T^*$ value, and finally only a broad Raman peak remains at higher temperatures (see Fig.~\ref{fig:3} and Fig.~S4).
This suggests that the fine structure of the A$_\textrm{g}^5$ mode can be observed not only in the pure ferromagnetic phase, but also is apparent in the intermediate phase.

The observed fine structure of a Raman mode has recently been reported by Chen~\cite{Chen}.
The Authors presented a model involving nonlinear interactions between optical phonons.
We have proposed an alternative explanation of the observed structure, which we relate to the isotope effects (for review see~\cite{Cardona}).

The calculated $\Gamma$-point eigenvector of the A$_\textrm{g}^5$ mode confirms that its vibrational character is strongly dominated by the Ge–Ge pair. As summarized in Table SII of the Supplementary Information, the displacement amplitude of each Ge atom is approximately 0.697, whereas the corresponding amplitudes are only about 0.068 for Te and 0.005 for Cr. Thus, the Ge displacement is approximately one order of magnitude larger than that of Te and more than two orders of magnitude larger than that of Cr. Moreover, the two Ge atoms move in opposite directions, consistent with a predominantly Ge-Ge stretching vibration. The substantially smaller displacements of the surrounding Cr and Te atoms indicate that their contribution to this mode is comparatively weak. Therefore, to a good approximation, the A$_\textrm{g}^5$ vibration can be treated as an effective diatomic vibration of the two Ge atoms, forming a Ge-Ge pseudo-molecule, as schematically illustrated in the inset of Fig. \ref{fig:5}. The complete Cartesian components and magnitudes of the calculated atomic displacements are provided in Table SII of the SI.

This results in an extremely flat dispersion of the A$_\textrm{g}^5$ mode throughout the BZ, as presented in Fig.~\ref{fig:2}.
We show that the observed fine structure of the A$_\textrm{g}^5$ mode can be explained by considering the five naturally occurring germanium isotopes, $i.e.$, 70, 72, 73, 74, and 76 with their corresponding abundances, 20.38$\%$, 27.31$\%$, 7.76$\%$, 36.72$\%$, and 7.83$\%$~\cite{deLaeter2003}.
Consequently, the Ge-Ge vibration in the A$_\textrm{g}^5$ mode can originate from Ge pairs formed by various 15 isotope configurations, $e.g.$ 70-70, 72-76, etc.
Using a simple approach with two single point masses connected by a spring, we extracted the relative frequency spectrum for different Ge molecules, because the frequency of a given isotope composition is inversely proportional to the square root of the Ge-Ge pair reduced mass; see the SI for details.
The intensity of a given vibration was taken as a product of the natural abundances of isotopes forming a Ge-Ge pair.
Finally, we simulate the Raman spectrum of the A$_\textrm{g}^5$ mode assuming the Lorentz distribution of a given peak with input of its calculated frequency and intensity and the determined experimental linewidth of 0.4~cm$^{-1}$, which is presented in Fig.~\ref{fig:5}.
The comparison of the measured experimentally and simulated theoretically of the A$_\textrm{g}^5$ fine structures is excellent.
In addition, we extracted the value of the force constant ($K$) of the Ge pseudo-molecule in the CGT crystal on the level of about 187.2~N/m using the model discussed in the SI. 
The found $K$ value is much smaller compared to about 2300~N/m and 1200~N/m reported correspondingly to the well-known diatomic molecules of N$_2$ and O$_2$~\cite{Empedocles1967}, but is similar to the one of Si$_2$ molecules ($\sim$70~N/m) and larger than the ones of Li$_2$ and Na$_2$ molecules ($\sim$30~N/m and $\sim$20~N/m, respectively)~\cite{Empedocles1967}.
Note that the theoretically predicted $K$ constant for the ground state of the Ge$_2$ molecule is calculated to be 270~N/m~\cite{Anderson1975}, which is two times larger than 187~N/m for the investigated Ge pseudo-molecule in the CGT crystal.

We note that a similar fine structure of a Raman scattering line was previously reported in natural Ge~\cite{Mayur}. 
The spectrum of Ge-O quasimolecules in natural Ge at 2.03~K showed the complex fine structure associated with each of the distinct isotopic combinations of the two Ge atoms in the Ge-O quasimolecule identified by the average isotopic mass of Ge for each pair of lines.
This seems to strongly support our explanation of the Ge-Ge stretching-mode fine structure in the investigated CGT. 

%XXXXXXXXXXXXXXXXXXXXXXXX        SUMMARY
\section{Summary \label{sec:Summary}}
In conclusion, we have conducted systematic RS investigations of the vibrational properties of CGT. 
All theoretically predicted phonon modes were observed at low temperatures ($T$=5~K), in agreement with the calculated phonon dispersion. 
Temperature-dependent Raman spectroscopy revealed two distinct phase transitions: a transition from the ferromagnetic to an intermediate phase at approximately 60~K, and a subsequent transition to the paramagnetic phase around 150~K. 
These phase transitions are reflected not only in the phonon energies but also in the linewidths and intensities of the Raman modes. 
We analyze the spin-phonon coupling responsible for the observed energy shifts. 
Furthermore, the observed A$_g^5$ fine structure has been successfully simulated using a model that takes into account vibrations of Ge-Ge pseudomolecules for various Ge isotopes.

%XXXXXXXXXXXXXXXXXXXXXXXX        EXPERIMENTAL METHODS
\section{Methods \label{sec:methods}}
To synthesize single CGT crystals, we employed a self-flux method, leveraging a germanium-tellurium rich flux, based on established procedures~\cite{Zeisner2019}. 
All precursor handling occurred within an M-Braun glovebox, maintained in a rigorously controlled argon atmosphere (H$_2$O and O$_2$ levels below 0.1 ppm). 
A mixture of chromium granules (4N purity, Thermo Fisher), germanium (4N purity, ChemPur), and tellurium lumps (5N purity, Alfa Aesar) was combined in a 10:13:77 molar ratio (Cr:Ge:Te) and loaded into an alumina crucible. 
This crucible was then sealed within a 19 mm diameter fused-silica tube under a partial argon atmosphere (approximately 400 mbar). 
The crystal growth process involved a carefully controlled thermal profile: the ampoule was heated to 1000$^{\circ}$C over 6 hours, kept at this temperature for 24 hours to ensure thorough mixing and dissolution, and subsequently cooled to 450$^{\circ}$C at a rate of 2$^{\circ}$C per hour. 
At 450$^{\circ}$C, the excess Ge-Te flux was decanted by centrifugation, revealing plate-like millimeter-sized CGT crystals. 
The crystal structure was determined using single-crystal X-ray diffraction, and the results, which align well with previously reported data~\cite{Carteaux1995, Yang2016}, are discussed in a different publication~\cite{Abadia-Huguet2025}.

The CGT flakes were placed on a Si/(300 nm)SiO$_2$ substrate by polydimethylsiloxane (PDMS)-based exfoliation~\cite{Gomez2014} of the bulk crystals. 
The PDMS stamp was prepared from the gel-film purchased from Gel-Pak. 
The flakes of interest were initially identified by visual inspection under an optical microscope and then subjected to
atomic force microscopy, confirming their thicknesses. 

The RS experiments were conducted using a series of laser excitations: $\lambda =$ 405 nm (3.06 eV), $\lambda =$ 515 nm (2.41 eV), $\lambda =$ 561 nm (2.21 eV), $\lambda =$ 633 nm (1.96 eV), and $\lambda =$ 785 nm (1.58 eV). 
The excitation power focused on the sample was kept at approx. 1~mW in all investigations.
The measurements were performed with excitation light focused using a 50× long-working-distance objective with a numerical aperture (NA) of 0.55, producing a spot diameter of approximately 1 $\mu$m. 
The signal was collected in backscattering geometry through the same microscope objective and dispersed by a spectrometer with a focal length of 0.75 m and 1800 g/mm holographic grating.  Finally, the light was collected using a liquid-nitrogen-cooled charge-coupled device (CCD) camera.
All measurements, including temperature-dependent experiments in the range from $T$=5~K to 300~K, were done by placing the sample on a cold finger in a continuous flow cryostat mounted on x–y motorized positioners.
Polarization-sensitive RS measurements were carried out in two configurations: co-linear (XX) and cross-linear (XY), corresponding to parallel and perpendicular orientations of the excitation and detection polarization axes, respectively. 
The RS signal analysis was employed using a motorized half-wave plate mounted in the detection path.
To measure low-energy Raman scattering (below 50 cm$^{-1}$ from the laser line), a set of Bragg-grating Notch filters was incorporated into the detection paths.

% DFT calculations
DFT calculations were performed in Vienna ab-initio simulation package (VASP) version 6.4.2~\cite{KRESSE199615, PhysRevB.54.11169}. 
The projector augment wave (PAW) potentials Cr$\_{pv}$, Ge$\_d$ and Te with Pedew-Burke-Ernzerhof (PBE) parametrization of the general gradients approximation (GGA) to the exchange-correlation functional were used \cite{ PhysRevB.59.1758}. 
The unit cell parameters and atomic positions were optimized with criteria of 10$^{-5}$ ~eV/Å for forces and 0.1 kbar for stresses, including the D3 correction to van der Waals interactions \cite{10.1063/1.3382344}. 
An energy cut-off of 400 eV and a 6$\times$6$\times$6 $\Gamma$-centered k-points grid were used. 
The GGA+U approximation with effective Hubbard parameter U=1~eV was employed to properly model the ferromagnetic order of the magnetic moments of Cr atoms~\cite{Kang2019}. 
Phonon calculations were performed within the direct method as implemented in Phonopy \cite{PhysRevLett.78.4063,phonopy-phono3py-JPSJ,phonopy-phono3py-JPCM}. A 2$\times$2$\times$2 supercell was used to evaluate the interatomic force constants.

\section*{Data availability}
The datasets generated during and/or analysed during the current study are available in the UW Research Data repository,  \url{http://doi.org/XXX}.

%XXXXXXXXXXXXXXXXXXXXXXXX        ACKNOWLEDGMENTS
\section*{Funding}
The work has been supported by the National Science Centre, Poland (Grant No. 2020/37/B/ST3/02311, 2023/48/C/ST3/00309, 2025/57/B/ST5/03288).
DFT calculations have been carried out using computational resources of the Interdisciplinary Centre for Mathematical and Computational Modelling (ICM), University of Warsaw, under grant GB83-21.

\section*{Author contributions}
G.K., K.O.-P., A.B., and M.R.M. performed the experiments. G.K. fabricated the samples with exfoliated flakes. T.W. carried out DFT calculations. M.I.S. and H.K. grew the CGT crystals. M.R.M. initiated and supervised the project. G.K., A.B., and M.R.M. wrote the manuscript with inputs from all co-authors.

\section*{Competing interests}
The authors declare no competing interests.

\bibliographystyle{apsrev4-2}
\bibliography{biblio}

@article{KRESSE199615,
title = {Efficiency of ab-initio total energy calculations for metals and semiconductors using a plane-wave basis set},
journal = {Computational Materials Science},
volume = {6},
number = {1},
pages = {15-50},
year = {1996},
issn = {0927-0256},
doi = {https://doi.org/10.1016/0927-0256(96)00008-0},
url = {https://www.sciencedirect.com/science/article/pii/0927025696000080},
author = {G. Kresse and J. Furthmüller}
}

@article{PhysRevB.54.11169,
  title = {Efficient iterative schemes for ab initio total-energy calculations using a plane-wave basis set},
  author = {Kresse, G. and Furthm\"uller, J.},
  journal = {Phys. Rev. B},
  volume = {54},
  issue = {16},
  pages = {11169--11186},
  numpages = {0},
  year = {1996},
  month = {Oct},
  publisher = {American Physical Society},
  doi = {10.1103/PhysRevB.54.11169},
  url = {https://link.aps.org/doi/10.1103/PhysRevB.54.11169}
}

@article{PhysRevB.59.1758,
  title = {From ultrasoft pseudopotentials to the projector augmented-wave method},
  author = {Kresse, G. and Joubert, D.},
  journal = {Phys. Rev. B},
  volume = {59},
  issue = {3},
  pages = {1758--1775},
  numpages = {0},
  year = {1999},
  month = {Jan},
  publisher = {American Physical Society},
  doi = {10.1103/PhysRevB.59.1758},
  url = {https://link.aps.org/doi/10.1103/PhysRevB.59.1758}
}

@article{10.1063/1.3382344,
    author = {Grimme, Stefan and Antony, Jens and Ehrlich, Stephan and Krieg, Helge},
    title = {A consistent and accurate ab initio parametrization of density functional dispersion correction (DFT-D) for the 94 elements H-Pu},
    journal = {The Journal of Chemical Physics},
    volume = {132},
    number = {15},
    pages = {154104},
    year = {2010},
    month = {04},
    issn = {0021-9606},
    doi = {10.1063/1.3382344},
    url = {https://doi.org/10.1063/1.3382344}
}

@Article{Kang2019,
author={Kang, Sungmo
and Kang, Seungjin
and Yu, Jaejun},
title={Effect of Coulomb Interactions on the Electronic and Magnetic Properties of Two-Dimensional CrSiTe3 and CrGeTe3 Materials},
journal={Journal of Electronic Materials},
year={2019},
month={Mar},
day={01},
volume={48},
number={3},
pages={1441-1445},
issn={1543-186X},
doi={10.1007/s11664-018-6601-2},
url={https://doi.org/10.1007/s11664-018-6601-2}
}

@article{phonopy-phono3py-JPCM,
  author  = {Togo, Atsushi and Chaput, Laurent and Tadano, Terumasa and Tanaka, Isao},
  title   = {Implementation strategies in phonopy and phono3py},
  journal = {J. Phys. Condens. Matter},
  volume  = {35},
  number  = {35},
  pages   = {353001},
  year    = {2023},
  doi     = {10.1088/1361-648X/acd831}
}

@article{phonopy-phono3py-JPSJ,
  author  = {Togo, Atsushi},
  title   = {First-principles Phonon Calculations with Phonopy and Phono3py},
  journal = {J. Phys. Soc. Jpn.},
  volume  = {92},
  number  = {1},
  pages   = {012001},
  year    = {2023},
  doi     = {10.7566/JPSJ.92.012001}
}

@article{PhysRevLett.78.4063,
  title = {First-Principles Determination of the Soft Mode in Cubic ${\mathrm{ZrO}}_{2}$},
  author = {Parlinski, K. and Li, Z. Q. and Kawazoe, Y.},
  journal = {Phys. Rev. Lett.},
  volume = {78},
  issue = {21},
  pages = {4063--4066},
  numpages = {0},
  year = {1997},
  month = {May},
  publisher = {American Physical Society},
  doi = {10.1103/PhysRevLett.78.4063},
  url = {https://link.aps.org/doi/10.1103/PhysRevLett.78.4063}
}

@article{Carteaux1995,
doi = {10.1088/0953-8984/7/1/008},
url = {https://dx.doi.org/10.1088/0953-8984/7/1/008},
year = {1995},
month = {jan},
publisher = {},
volume = {7},
number = {1},
pages = {69},
author = {V Carteaux and D Brunet and G Ouvrard and G Andre},
title = {Crystallographic, magnetic and electronic structures of a new layered ferromagnetic compound Cr2Ge2Te6},
journal = {Journal of Physics: Condensed Matter}
}

@article{Ji2013,
    author = {Ji, Huiwen and Stokes, R. A. and Alegria, L. D. and Blomberg, E. C. and Tanatar, M. A. and Reijnders, Anjan and Schoop, L. M. and Liang, Tian and Prozorov, R. and Burch, K. S. and Ong, N. P. and Petta, J. R. and Cava, R. J.},
    title = {A ferromagnetic insulating substrate for the epitaxial growth of topological insulators},
    journal = {Journal of Applied Physics},
    volume = {114},
    number = {11},
    pages = {114907},
    year = {2013},
    month = {09},
    issn = {0021-8979},
    doi = {10.1063/1.4822092},
    url = {https://doi.org/10.1063/1.4822092}
}

@Article{Gong2017,
author={Gong, Cheng and Li, Lin and Li, Zhenglu and Ji, Huiwen
and Stern, Alex and Xia, Yang and Cao, Ting and Bao, Wei
and Wang, Chenzhe and Wang, Yuan and Qiu, Z. Q. and Cava, R. J.
and Louie, Steven G. and Xia, Jing and Zhang, Xiang},
title={Discovery of intrinsic ferromagnetism in two-dimensional van der Waals crystals},
journal={Nature},
year={2017},
month={Jun},
day={01},
volume={546},
number={7657},
pages={265-269},
issn={1476-4687},
doi={10.1038/nature22060},
url={https://doi.org/10.1038/nature22060}
}

@Article{Zhang2016,
author={Zhang, Xiao and Zhao, Yuelei and Song, Qi and Jia, Shuang and Shi, Jing and Han, Wei},
title={Magnetic anisotropy of the single-crystalline ferromagnetic insulator Cr2Ge2Te6},
journal={Japanese Journal of Applied Physics},
year={2016},
month={Feb},
day={01},
publisher={The Japan Society of Applied Physics},
volume={55},
number={3},
pages={033001},
issn={1347-4065},
doi={10.7567/JJAP.55.033001},
url={https://dx.doi.org/10.7567/JJAP.55.033001}
}

@Article{Siskins2022,
author={{\v{S}}i{\v{s}}kins, Makars and Kurdi, Samer and Lee, Martin and Slotboom, Benjamin J. M. and Xing, Wenyu and Ma{\~{n}}as-Valero, Samuel and Coronado, Eugenio and Jia, Shuang and Han, Wei and van der Sar, Toeno and van der Zant, Herre S. J. and Steeneken, Peter G.},
title={Nanomechanical probing and strain tuning of the Curie temperature in suspended Cr2Ge2Te6-based heterostructures},
journal={npj 2D Materials and Applications},
year={2022},
month={Jun},
day={15},
volume={6},
number={1},
pages={41},
issn={2397-7132},
doi={10.1038/s41699-022-00315-7},
url={https://doi.org/10.1038/s41699-022-00315-7}
}

@Article{Verzhbitskiy2020,
author={Verzhbitskiy, Ivan A. and Kurebayashi, Hidekazu and Cheng, Haixia and Zhou, Jun and Khan, Safe and Feng, Yuan Ping and Eda, Goki},
title={Controlling the magnetic anisotropy in Cr2Ge2Te6 by electrostatic gating},
journal={Nature Electronics},
year={2020},
month={Aug},
day={01},
volume={3},
number={8},
pages={460-465},
issn={2520-1131},
doi={10.1038/s41928-020-0427-7},
url={https://doi.org/10.1038/s41928-020-0427-7}
}

@Article{Xing2017,
author={Xing, Wenyu and Chen, Yangyang and Odenthal, Patrick M. and Zhang, Xiao and Yuan, Wei and Su, Tang and Song, Qi and Wang, Tianyu and Zhong, Jiangnan and Jia, Shuang and Xie, X. C. and Li, Yan and Han, Wei},
title={Electric field effect in multilayer Cr2Ge2Te6: a ferromagnetic 2D material},
journal={2D Materials},
year={2017},
month={May},
day={22},
publisher={IOP Publishing},
volume={4},
number={2},
pages={024009},
issn={2053-1583},
doi={10.1088/2053-1583/aa7034},
url={https://dx.doi.org/10.1088/2053-1583/aa7034}
}

@Article{Zhuo2021,
author={Zhuo, Weizhuang and Lei, Bin and Wu, Shuang and Yu, Fanghang and Zhu, Changsheng and Cui, Jianhua and Sun, Zeliang and Ma, Donghui and Shi, Mengzhu and Wang, Honghui and Wang, Wenxiang and Wu, Tao and Ying, Jianjun and Wu, Shiwei and Wang, Zhenyu and Chen, Xianhui},
title={Manipulating Ferromagnetism in Few-Layered Cr2Ge2Te6},
journal={Advanced Materials},
year={2021},
month={Aug},
day={01},
publisher={John Wiley {\&} Sons, Ltd},
volume={33},
number={31},
pages={2008586},
issn={0935-9648},
doi={10.1002/adma.202008586},
url={https://doi.org/10.1002/adma.202008586}
}

@article{Lin2018,
  title = {Pressure-induced spin reorientation transition in layered ferromagnetic insulator ${\mathrm{Cr}}_{2}{\mathrm{Ge}}_{2}{\mathrm{Te}}_{6}$},
  author = {Lin, Zhisheng and Lohmann, Mark and Ali, Zulfikhar A. and Tang, Chi and Li, Junxue and Xing, Wenyu and Zhong, Jiangnan and Jia, Shuang and Han, Wei and Coh, Sinisa and Beyermann, Ward and Shi, Jing},
  journal = {Phys. Rev. Mater.},
  volume = {2},
  issue = {5},
  pages = {051004},
  numpages = {6},
  year = {2018},
  month = {May},
  publisher = {American Physical Society},
  doi = {10.1103/PhysRevMaterials.2.051004},
  url = {https://link.aps.org/doi/10.1103/PhysRevMaterials.2.051004}
}

@Article{Sun2018,
author={Sun, Y. and Xiao, R. C. and Lin, G. T. and Zhang, R. R. and Ling, L. S. and Ma, Z. W. and Luo, X. and Lu, W. J. and Sun, Y. P. and Sheng, Z. G.},
title={Effects of hydrostatic pressure on spin-lattice coupling in two-dimensional ferromagnetic Cr2Ge2Te6},
journal={Applied Physics Letters},
year={2018},
month={Feb},
day={16},
volume={112},
number={7},
pages={072409},
issn={0003-6951},
doi={10.1063/1.5016568},
url={https://doi.org/10.1063/1.5016568}
}

@Article{Tian2016,
author={Tian, Yao and Gray, Mason J. and Ji, Huiwen and Cava, R. J. and Burch, Kenneth S.},
title={Magneto-elastic coupling in a potential ferromagnetic 2D atomic crystal},
journal={2D Materials},
year={2016},
month={Jun},
day={23},
publisher={IOP Publishing},
volume={3},
number={2},
pages={025035},
issn={2053-1583},
doi={10.1088/2053-1583/3/2/025035},
url={https://dx.doi.org/10.1088/2053-1583/3/2/025035}
}

@Article{Samanta2024,
author={Samanta, Sudeshna and Iturriaga, Hector and Mai, Thuc T. and Biacchi, Adam J. and Islam, Rajibul and Fullerton, John and Hight Walker, Angela R. and Noufal, Mohamed and Siebenaller, Ryan and Rowe, Emmanuel and Phatak, Charudatta and Susner, Michael A. and Xue, Fei and Singamaneni, Srinivasa R.},
title={Spin--Phonon Coupling and Magnetic Transition in an Organic Molecule Intercalated Cr2Ge2Te6},
journal={Nano Letters},
year={2024},
month={Jul},
day={31},
publisher={American Chemical Society},
volume={24},
number={30},
pages={9169-9177},
issn={1530-6984},
doi={10.1021/acs.nanolett.4c00976},
url={https://doi.org/10.1021/acs.nanolett.4c00976}
}

@Article{Huang2024,
author={Huang, Zhengjie and McCray, Arthur R. C. and Li, Yue and Morrow, Darien J. and Qian, Eric K. and Young Chung, Duck and Kanatzidis, Mercouri G. and Phatak, Charudatta and Ma, Xuedan},
title={Raman Shifts in Two-Dimensional van der Waals Magnets Reveal Magnetic Texture Evolution},
journal={Nano Letters},
year={2024},
month={Feb},
day={07},
publisher={American Chemical Society},
volume={24},
number={5},
pages={1531-1538},
issn={1530-6984},
doi={10.1021/acs.nanolett.3c03923},
url={https://doi.org/10.1021/acs.nanolett.3c03923}
}

@article{Idzuchi2025,
  title = {Enhanced ferromagnetism in an artificially stretched lattice in quasi-two-dimensional $\mathrm{C}{\mathrm{r}}_{2}\mathrm{G}{\mathrm{e}}_{2}\mathrm{T}{\mathrm{e}}_{6}$},
  author = {Idzuchi, Hiroshi and Llacsahuanga Allcca, Andres E. and Lu, Anh Khoa Augustin and Saito, Mitsuhiro and Das, Subhadip and Ribeiro, John F. and Houssa, Michel and Meng, Ruishen and Inoue, Kazutoshi and Pan, Xing-Chen and Tanigaki, Katsumi and Ikuhara, Yuichi and Nakanishi, Takeshi and Chen, Yong P.},
  journal = {Phys. Rev. B},
  volume = {111},
  issue = {2},
  pages = {L020402},
  numpages = {7},
  year = {2025},
  month = {Jan},
  publisher = {American Physical Society},
  doi = {10.1103/PhysRevB.111.L020402},
  url = {https://link.aps.org/doi/10.1103/PhysRevB.111.L020402}
}

@article{Chakkar2024,
  title = {Broken weak and strong spin rotational symmetries and tunable interactions between phonons and the continuum in $\mathrm{C}{\mathrm{r}}_{2}\mathrm{G}{\mathrm{e}}_{2}\mathrm{T}{\mathrm{e}}_{6}$},
  author = {Chakkar, Atul G. and Kumar, Deepu and Kumar, Pradeep},
  journal = {Phys. Rev. B},
  volume = {109},
  issue = {13},
  pages = {134406},
  numpages = {11},
  year = {2024},
  month = {Apr},
  publisher = {American Physical Society},
  doi = {10.1103/PhysRevB.109.134406},
  url = {https://link.aps.org/doi/10.1103/PhysRevB.109.134406}
}

@Article{Dong2020,
author={Dong, Enlai and Liu, Bo and Dong, Qing and Shi, Xuhan and Ma, Xin and Liu, Ran and Zhu, Xuebin and Luo, Xuan and Li, Xiaodong and Li, Yanchun and Li, Quanjun and Liu, Bingbing},
title={Effects of pressure on the structure and properties of layered ferromagnetic Cr2Ge2Te6},
journal={Physica B: Condensed Matter},
year={2020},
month={Oct},
day={15},
volume={595},
pages={412344},
issn={0921-4526},
url={https://www.sciencedirect.com/science/article/pii/S0921452620303550}
}

@article{Zeisner2019,
  title = {Magnetic anisotropy and spin-polarized two-dimensional electron gas in the van der Waals ferromagnet ${\mathrm{Cr}}_{2}{\mathrm{Ge}}_{2}{\mathrm{Te}}_{6}$},
  author = {Zeisner, J. and Alfonsov, A. and Selter, S. and Aswartham, S. and Ghimire, M. P. and Richter, M. and van den Brink, J. and B\"uchner, B. and Kataev, V.},
  journal = {Phys. Rev. B},
  volume = {99},
  issue = {16},
  pages = {165109},
  numpages = {14},
  year = {2019},
  month = {Apr},
  publisher = {American Physical Society},
  doi = {10.1103/PhysRevB.99.165109},
  url = {https://link.aps.org/doi/10.1103/PhysRevB.99.165109}
}

@Article{Yang2016,
author={Yang, Dingfeng and Yao, Wei and Chen, Qiufan and Peng, Kunling and Jiang, Pengfei and Lu, Xu and Uher, Ctirad and Yang, Tao and Wang, Guoyu and Zhou, Xiaoyuan},
title={Cr2Ge2Te6: High Thermoelectric Performance from Layered Structure with High Symmetry},
journal={Chemistry of Materials},
year={2016},
month={Mar},
day={22},
publisher={American Chemical Society},
volume={28},
number={6},
pages={1611-1615},
issn={0897-4756},
doi={10.1021/acs.chemmater.5b04895},
url={https://doi.org/10.1021/acs.chemmater.5b04895}
}

@Article{Abadia-Huguet2025,
author={Abadia-Huguet, Aleix and Mendive-Tapia, Eduardo and Stern-Taulats, Enric and Planes, Antoni and Eggert, Benedikt and Wende, Heiko and Acet, Mehmet and Sturza, Mihai-Ionut and Kohlmann, Holger and Costache, Marius V. and Ma{\~{n}}osa, Llu{\'i}s},
title={Multicaloric effects and magnetostructural coupling in the Cr2Ge2Te6 van der Waals crystal},
journal={Applied Materials Today},
year={2025},
month={Jun},
day={01},
volume={44},
pages={102749},
issn={2352-9407},
url={https://www.sciencedirect.com/science/article/pii/S2352940725001684}
}

@article{Gomez2014,
doi = {10.1088/2053-1583/1/1/011002},
url = {https://dx.doi.org/10.1088/2053-1583/1/1/011002},
year = {2014},
month = {apr},
publisher = {IOP Publishing},
volume = {1},
number = {1},
pages = {011002},
author = {Castellanos-Gomez, Andres and Buscema, Michele and Molenaar, Rianda and Singh, Vibhor and Janssen, Laurens and van der Zant, Herre S J and Steele, Gary A},
title = {Deterministic transfer of two-dimensional materials by all-dry viscoelastic stamping},
journal = {2D Materials}
}

@article{Balkanski1983,
  title = {Anharmonic effects in light scattering due to optical phonons in silicon},
  author = {Balkanski, M. and Wallis, R. F. and Haro, E.},
  journal = {Phys. Rev. B},
  volume = {28},
  issue = {4},
  pages = {1928--1934},
  numpages = {0},
  year = {1983},
  month = {Aug},
  publisher = {American Physical Society},
  doi = {10.1103/PhysRevB.28.1928},
  url = {https://link.aps.org/doi/10.1103/PhysRevB.28.1928}
}

@Article{Muhammad2024,
author={Muhammad, Zahir and Hussain, Ghulam and Islam, Rajbul and Zawadzka, Natalia and Hossain, Md Shafayat and Iqbal, Obaid and Babi{\'{n}}ski, Adam and Molas, Maciej R. and Xue, Fei and Zhang, Yue and Hasan, M. Zahid and Zhao, Weisheng},
title={Electronic Transport and Interaction of Lattice Dynamics in Topological Nodalline Semimetal HfAs2 Single Crystals},
journal={Advanced Functional Materials},
year={2024},
month={Oct},
day={01},
publisher={John Wiley {\&} Sons, Ltd},
volume={34},
number={41},
pages={2316775},
issn={1616-301X},
doi={10.1002/adfm.202316775},
url={https://doi.org/10.1002/adfm.202316775}
}

@Article{Buruiana2022,
author={Buruiana, A. T. and Bocirnea, A. E. and Kuncser, A. C.
and Tite, T. and Matei, E. and Mihai, C. and Zawadzka, N. and Olkowska-Pucko, K. and Kipczak, {\L} and Babi{\'{n}}ski, A. and Molas, M. R. and Velea, A. and Galca, A. C.},
title={Layered SnSe nanoflakes with anharmonic phonon properties and memristive characteristics},
journal={Applied Surface Science},
year={2022},
month={Oct},
day={15},
volume={599},
pages={153983},
issn={0169-4332},
url={https://www.sciencedirect.com/science/article/pii/S0169433222015252}
}

@Article{Lapinska2016,
  author  = {Łapińska, Anna and Taube, Andrzej and Judek, Jaroslaw and Zdrojek, Mariusz},
  journal = {The Journal of Physical Chemistry C},
  title   = {Temperature Evolution of Phonon Properties in Few-Layer Black Phosphorus},
  year    = {2016},
  number  = {9},
  pages   = {5265-5270},
  volume  = {120},
  doi     = {10.1021/acs.jpcc.6b01468},
  url     = {https://doi.org/10.1021/acs.jpcc.6b01468}
}

@article{Han2019,
author = {Han, Myung-Geun and Garlow, Joseph A. and Liu, Yu and Zhang, Huiqin and Li, Jun and DiMarzio, Donald and Knight, Mark W. and Petrovic, Cedomir and Jariwala, Deep and Zhu, Yimei},
title = {Topological Magnetic-Spin Textures in Two-Dimensional van der Waals Cr2Ge2Te6},
journal = {Nano Letters},
volume = {19},
number = {11},
pages = {7859-7865},
year = {2019},
doi = {10.1021/acs.nanolett.9b02849},
URL = {https://doi.org/10.1021/acs.nanolett.9b02849}
}

@article{Koo2022,
author = {Koo, Hyun-Joo and Kremer, Reinhard K. and Whangbo, Myung-Hwan},
title = {High-Spin Orbital Interactions Across van der Waals Gaps Controlling the Interlayer Ferromagnetism in van der Waals Ferromagnets},
journal = {Journal of the American Chemical Society},
volume = {144},
number = {36},
pages = {16272-16275},
year = {2022},
doi = {10.1021/jacs.2c06741},
URL = { https://doi.org/10.1021/jacs.2c06741}
}

@Article{Kozlenko2021,
author={Kozlenko, D. P. and Lis, O. N. and Kichanov, S. E. and Lukin, E. V. and Belozerova, N. M. and Savenko, B. N.},
title={Spin-induced negative thermal expansion and spin--phonon coupling in van der Waals material CrBr3},
journal={npj Quantum Materials},
year={2021},
month={Mar},
day={02},
volume={6},
number={1},
pages={19},
issn={2397-4648},
doi={10.1038/s41535-021-00318-5},
url={https://doi.org/10.1038/s41535-021-00318-5}
}

@article{Pawbake2023,
  title = {Raman scattering signatures of strong spin-phonon coupling in the bulk magnetic van der Waals material CrSBr},
  author = {Pawbake, Amit and Pelini, Thomas and Wilson, Nathan P. and Mosina, Kseniia and Sofer, Zdenek and Heid, Rolf and Faugeras, Clement},
  journal = {Phys. Rev. B},
  volume = {107},
  issue = {7},
  pages = {075421},
  numpages = {7},
  year = {2023},
  month = {Feb},
  publisher = {American Physical Society},
  doi = {10.1103/PhysRevB.107.075421},
  url = {https://link.aps.org/doi/10.1103/PhysRevB.107.075421}
}

@Article{Kipczak2024,
author={Kipczak, {\L}ucja and Karmakar, Arka and Grzeszczyk, Magdalena and Janiszewska, R{\'o}{\.{z}}a and Wo{\'{z}}niak, Tomasz and Chen, Zhaolong and Paw{\l}owski, Jan and Watanabe, Kenji and Taniguchi, Takashi and Babi{\'{n}}ski, Adam and Koperski, Maciej and Molas, Maciej R.},
title={Resonant Raman scattering of few layers CrBr3},
journal={Scientific Reports},
year={2024},
month={Mar},
day={29},
volume={14},
number={1},
pages={7484},
issn={2045-2322},
doi={10.1038/s41598-024-57622-w},
url={https://doi.org/10.1038/s41598-024-57622-w}
}

@Article{Froehlicher2015,
author={Froehlicher, Guillaume
and Lorchat, Etienne
and Fernique, Fran{\c{c}}ois
and Joshi, Chaitanya
and Molina-S{\'a}nchez, Alejandro
and Wirtz, Ludger
and Berciaud, St{\'e}phane},
title={Unified Description of the Optical Phonon Modes in N-Layer MoTe2},
journal={Nano Letters},
year={2015},
month={Oct},
day={14},
publisher={American Chemical Society},
volume={15},
number={10},
pages={6481-6489},
issn={1530-6984},
doi={10.1021/acs.nanolett.5b02683},
url={https://doi.org/10.1021/acs.nanolett.5b02683}
}

@Article{Grzeszczyk2016,
author={Grzeszczyk, M.
and Go{\l}asa, K.
and Zinkiewicz, M.
and Nogajewski, K.
and Molas, M. R.
and Potemski, M.
and Wysmo{\l}ek, A.
and Babi{\'{n}}ski, A.},
title={Raman scattering of few-layers MoTe2},
journal={2D Materials},
year={2016},
month={Apr},
day={13},
publisher={IOP Publishing},
volume={3},
number={2},
pages={025010},
issn={2053-1583},
doi={10.1088/2053-1583/3/2/025010},
url={https://dx.doi.org/10.1088/2053-1583/3/2/025010},
url={https://doi.org/10.1088/2053-1583/3/2/025010}
}

@Article{elahi2022,
  title={A review on two-dimensional (2D) magnetic materials and their potential applications in spintronics and spin-caloritronic},
  author={Elahi, Ehsan and Dastgeer, Ghulam and Nazir, Ghazanfar and Nisar, Sobia and Bashir, Mudasar and Qureshi, Haroon Akhter and Kim, Deok-kee and Aziz, Jamal and Aslam, Muhammad and Hussain, Kashif and others},
  journal={Computational Materials Science},
  volume={213},
  pages={111670},
  year={2022},
  publisher={Elsevier},
  url={https://doi.org/10.1016/j.commatsci.2022.111670}
}

@article{hoffmann2015,
  title={Opportunities at the frontiers of spintronics},
  author={Hoffmann, Axel and Bader, Sam D},
  journal={Physical Review Applied},
  volume={4},
  number={4},
  pages={047001},
  year={2015},
  publisher={APS},
  url={https://doi.org/10.1103/PhysRevApplied.4.047001}
}

@Article{Kipczak2020,
author={Kipczak, {\L}
and Grzeszczyk, M.
and Olkowska-Pucko, K.
and Babi{\'{n}}ski, A.
and Molas, M. R.},
title={The optical signature of few-layer ReSe2},
journal={Journal of Applied Physics},
year={2020},
month={Jul},
day={30},
volume={128},
number={4},
pages={044302},
issn={0021-8979},
doi={10.1063/5.0015289},
url={https://doi.org/10.1063/5.0015289}
}

@Article{deLaeter2003,
author={de Laeter, John R.
and B{\"o}hlke, John Karl
and De Bi{\`e}vre, P.
and Hidaka, H.
and Peiser, H. S.
and Rosman, K. J. R.
and Taylor, P. D. P.},
title={Atomic weights of the elements. Review 2000 (IUPAC Technical Report)},
year={2003},
month={2025},
journal = {UPAC Technical Report},
day={09-25},
volume={75},
number={6},
pages={683-800},
doi={10.1351/pac200375060683},
url={https://doi.org/10.1351/pac200375060683}
}

@Article{Gibertini2019,
author={Gibertini, M.
and Koperski, M.
and Morpurgo, A. F.
and Novoselov, K. S.},
title={Magnetic 2D materials and heterostructures},
journal={Nature Nanotechnology},
year={2019},
month={May},
day={01},
volume={14},
number={5},
pages={408-419},
issn={1748-3395},
doi={10.1038/s41565-019-0438-6},
url={https://doi.org/10.1038/s41565-019-0438-6}
}

@Article{Empedocles1967,
author={Empedocles, Philip},
title={Force Constants of Diatomic Molecules. I. A General Semitheoretical Approach},
journal={The Journal of Chemical Physics},
year={1967},
month={Jun},
day={01},
volume={46},
number={11},
pages={4474-4481},
issn={0021-9606},
doi={10.1063/1.1840571},
url={https://doi.org/10.1063/1.1840571}
}

@Article{Anderson1975,
author={Anderson, Alfred B.},
title={Vibrational potentials and structures in molecular and solid carbon, silicon, germanium, and tin},
journal={The Journal of Chemical Physics},
year={1975},
month={Nov},
day={15},
volume={63},
number={10},
pages={4430-4436},
issn={0021-9606},
doi={10.1063/1.431162},
url={https://doi.org/10.1063/1.431162}
}

@article{Chen,
  title = {Spontaneously formed phonon frequency combs in van der Waals solid {CrGeTe3} and {CrSiTe3}},
  author = {Chen, L. and Ye, G. and Nnokwe, C. and Pan, X.-C. and Tanigaki, K. and Cheng, G. and Chen, Y. P. and Yan, J. and Mandrus, David G. and Llacsahuanga Allcca, Andres E. and Giles-Donovan, Nathan and Birgeneau, Robert J. and He, Rui},
  journal = {Nat. Commun.},
  volume = {16},
  pages = {5795},
  year = {2025},
  doi = {10.1038/s41467-025-61173-7},
  url = {https://doi.org/10.1038/s41467-025-61173-7}
}

@article{Cardona,
  title = {Isotope effects on the optical spectra of semiconductors},
  author = {Cardona, Manuel and Thewalt, M. L. W.},
  journal = {Rev. Mod. Phys.},
  volume = {77},
  issue = {4},
  pages = {1173--1224},
  numpages = {0},
  year = {2005},
  month = {Nov},
  publisher = {American Physical Society},
  doi = {10.1103/RevModPhys.77.1173},
  url = {https://link.aps.org/doi/10.1103/RevModPhys.77.1173}
}

@article{Mayur,
  title = {Fine structure of the asymmetric stretching vibration of dispersed oxygen in monoisotopic germanium},
  author = {Mayur, A. J. and Sciacca, M. Dean and Udo, M. K. and Ramdas, A. K. and Itoh, K. and Wolk, J. and Haller, E. E.},
  journal = {Phys. Rev. B},
  volume = {49},
  issue = {23},
  pages = {16293--16299},
  numpages = {0},
  year = {1994},
  month = {Jun},
  publisher = {American Physical Society},
  doi = {10.1103/PhysRevB.49.16293},
  url = {https://link.aps.org/doi/10.1103/PhysRevB.49.16293}
}

@article{Singh2019_rev,
  title={Correlated paramagnetism and interplay of magnetic and phononic degrees of freedom in 3 d-5 d coupled La2CuIrO6},
  author={Singh, Birender and Kumar, Deepu and Manna, Kaustuv and Bera, AK and Cansever, G Aslan and Maljuk, A and Wurmehl, S and B{\"u}chner, B and Kumar, Pradeep},
  journal={Journal of Physics: Condensed Matter},
  volume={31},
  number={48},
  pages={485803},
  year={2019},
  publisher={IOP Publishing}
}

@Article{Spachmann2022_rev,
author={Spachmann, S.
and Elghandour, A.
and Selter, S.
and B{\"u}chner, B.
and Aswartham, S.
and Klingeler, R.},
journal={Physical Review Research},
year={2022},
month={May},
day={20},
publisher={American Physical Society},
volume={4},
number={2},
pages={L022040},
doi={10.1103/PhysRevResearch.4.L022040},
url={https://doi.org/10.1103/PhysRevResearch.4.L022040}
}

\newpage
\onecolumngrid

\renewcommand{\thefigure}{S\arabic{figure}}
\renewcommand{\thesubsection}{S\arabic{subsection}}
\renewcommand{\thetable}{S\Roman{table}}
\renewcommand{\theequation}{S\arabic{equation}}

\begin{center}
	%%%%%%%%% ABSTRACT TITLE
	{\large{ {\bf Supplementary Information: \\Spin-phonon coupling and isotope-related pseudo-molecule vibrations \\ in layered Cr$_2$Ge$_2$Te$_6$ ferromagnet}}}
	%%%%%%%%% ABSTRACT AUTHORS
	\vskip0.5\baselineskip{Grzegorz Krasucki,{$^{1}$} Katarzyna Olkowska-Pucko,{$^{1}$} Tomasz Woźniak,{$^{1}$} Mihai I. Sturza,{$^{2}$} Holger Kohlmann,{$^{2}$} Adam Babiński,{$^{1}$} and Maciej R. Molas{$^{1}$}}
	
	%%%%%%%%% AFFILIATION
	\vskip0.5\baselineskip{\em$^{1}$ University of Warsaw, Faculty of Physics, 02-093 Warsaw, Poland \\$^{2}$ Institute for Inorganic Chemistry, Leipzig University, D-04103 Leipzig, Germany}
	
\end{center}

\subsection{Optical and atomic force microscopy}

\begin{figure}[!th]
		\subfloat{}%
		\centering
		\includegraphics[width=0.9 \linewidth]{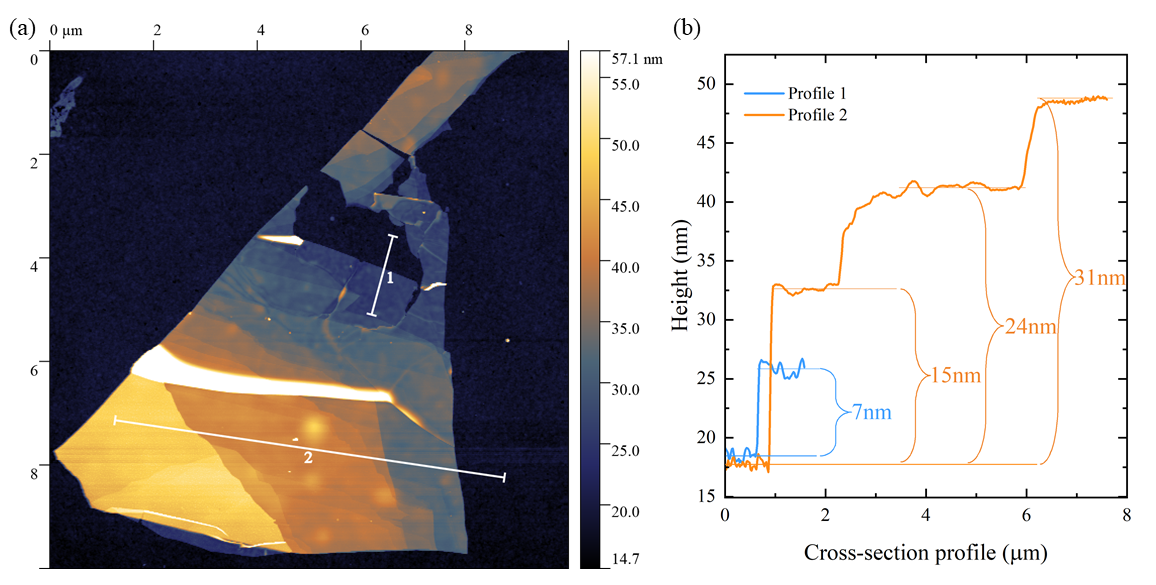}
    	\caption{(a) Atomic force microscopy topography image measured on the exfoliated CGT flake, with two profiles. (b) Cross-section data cut from AFM map, alongside marked profiles. Light-blue line corresponds to profile 1 and orange line to profile 2. On the plot it is marked corresponding heights of layers.}
		\label{AFM}
\end{figure}

Figure~\ref{AFM} (a) presents an atomic force microscopy (AFM) scan of the measured flake. 
This particular flake was used in the thickness-dependent study described in Section S5 of the Supplementary Information. 
Distinct terrace-like features can be clearly observed on the flake surface.
To extract corresponding terrace heights, cross-section data profiles are shown in Fig.~\ref{AFM}(b). 
Although flakes as thin as 7 nm were successfully identified, it was extremely challenging to isolate thinner flakes with lateral dimensions exceeding 1 $\mu$m$^2$. 
Most of the exfoliated flakes exhibited thicknesses comparable to those of the second level of the main terrace, approximately 24 nm in height. 
Consequently, temperature- and polarization-dependent measurements were performed on a flake of this representative thickness.

\newpage
\subsection{Excitation energy influence on the CGT Raman spectra}

\begin{figure}[!th]
		\subfloat{}%
		\centering
		\includegraphics[width=0.7 \linewidth]{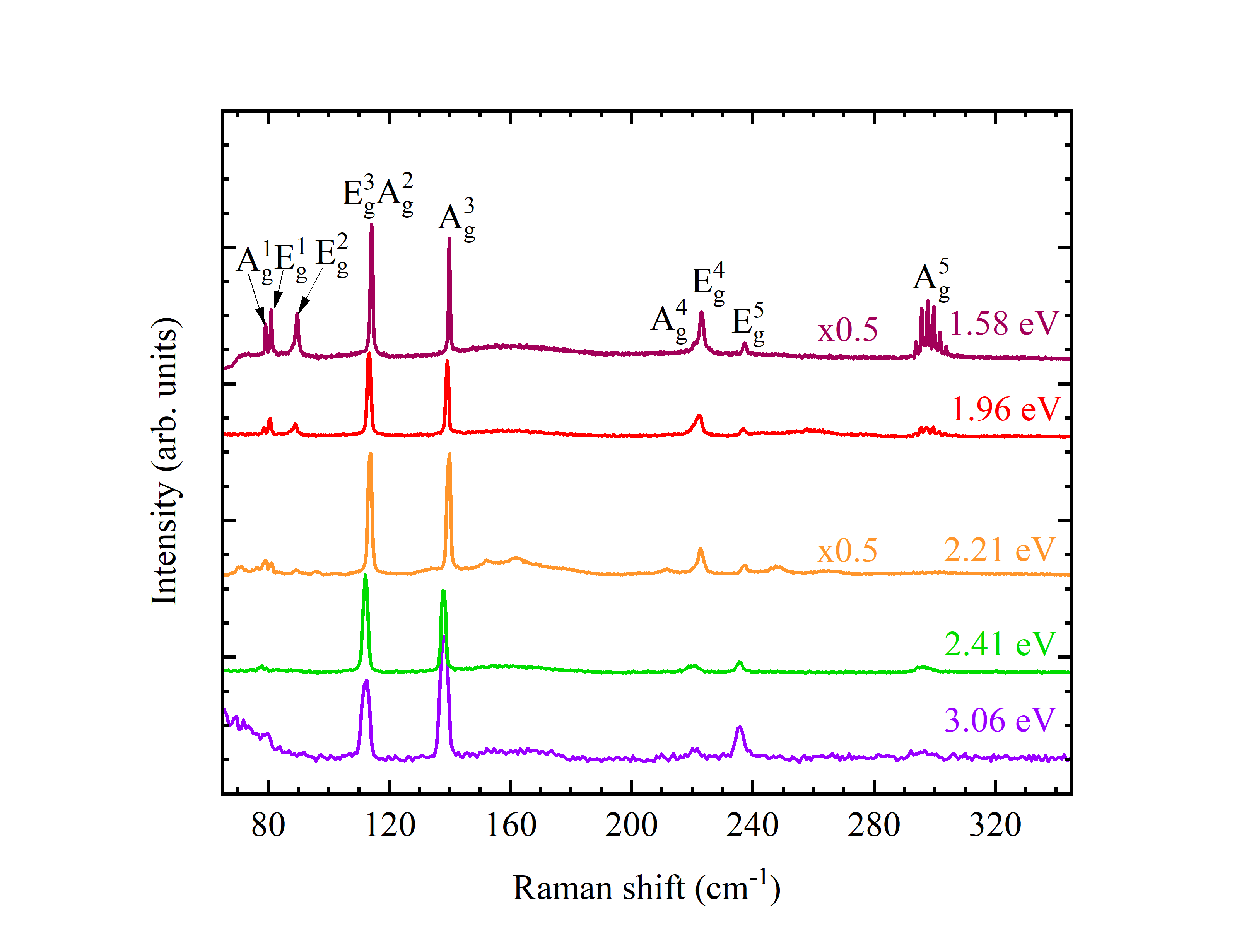}
    	\caption{Raman Scattering spectra of the exfoliated 24-nm thick CGT flake measured at 5~K with  different excitation energies: 1.58~eV, 1.96~eV, 2.21~eV, 2.41~eV, and 3.06~eV, using excitation power 1~mW. 
        The spectra are vertically shifted for clarity and some of them are multiplied by scaling factors to make them better visible.}
		\label{All_lasers}
\end{figure}

The representative low temperature ($T$=5~K) unpolarized Raman scattering (RS) spectra of Cr$_2$Ge$_2$Te$_6$ (CGT) crystals are presented in Fig.~\ref{All_lasers}.
We comparatively inspect the intensity dependence of the Raman peaks on the laser energy, utilizing 1.58~eV, 1.96~eV, 2.21~eV, 2.41~eV, and 3.06~eV excitations. 
Note that the RS spectra were normalized to the theoretical sensitivity of the experimental setup used, $i.e.$ reflection or transmission of optical elements (a beam splitter, mirrors, a lens, a long-pass filter), dispersion sensitivity of a grating in the spectrometer, and the efficiency of the CCD camera.
The number of observed phonon peaks and their intensities in the RS spectra of CGT are substantially affected by the excitation energy used.
For the 2.41~eV and 3.06~eV illuminations, only seven Raman modes can be easily resolved, while the 1.58~eV and 1.96~eV excitations allows us to unveil all the theoretically predicted phonon peaks, $i.e.$ 10.
The main difference between these two groups of the energies used originates from the increased spectral resolution of the experimental setup with the decreased excitation energy.
In particular, the fine structure of the A$_g^5$ mode, apparent as a set of discrete narrow lines, can be observed and resolved only for the 1.58~eV and 1.96~eV lasers.
Consequently, we decided to perform the following analysis of the phonon modes in the CGT crystal under the 1.58~eV illumination due to the highest possible spectral resolution of the Raman spectrum accompanied by its high intensity.

\newpage
\subsection{Polarization dependence of the CGT Raman spectra}

\begin{figure*}[!th]
		\subfloat{}%
		\centering
        \centering
    \parbox{.58\linewidth}{
        \centering
        \includegraphics[height=100mm]{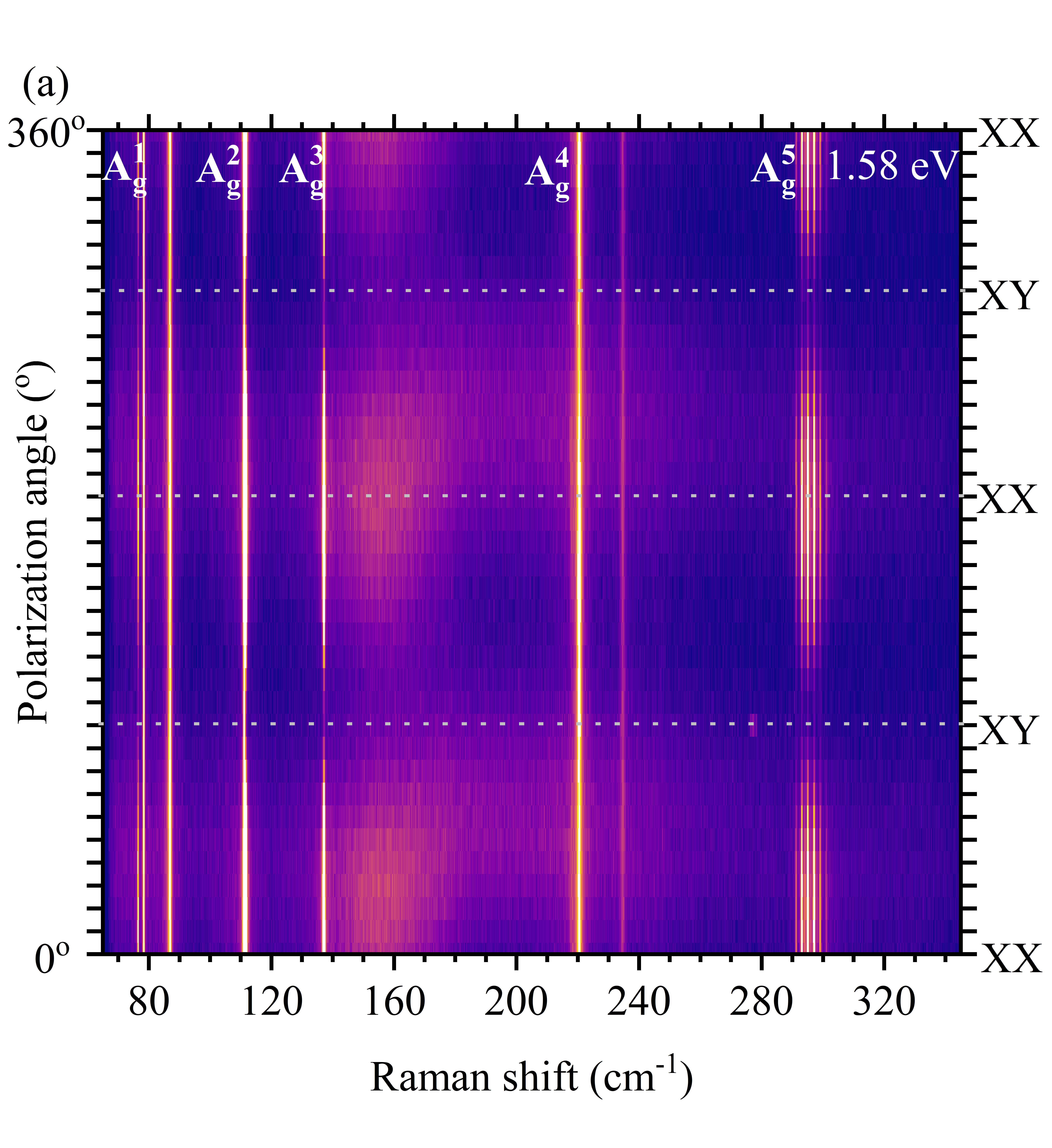}
         }
    \parbox{.38\linewidth}{
        \centering
         \includegraphics[height=100mm]{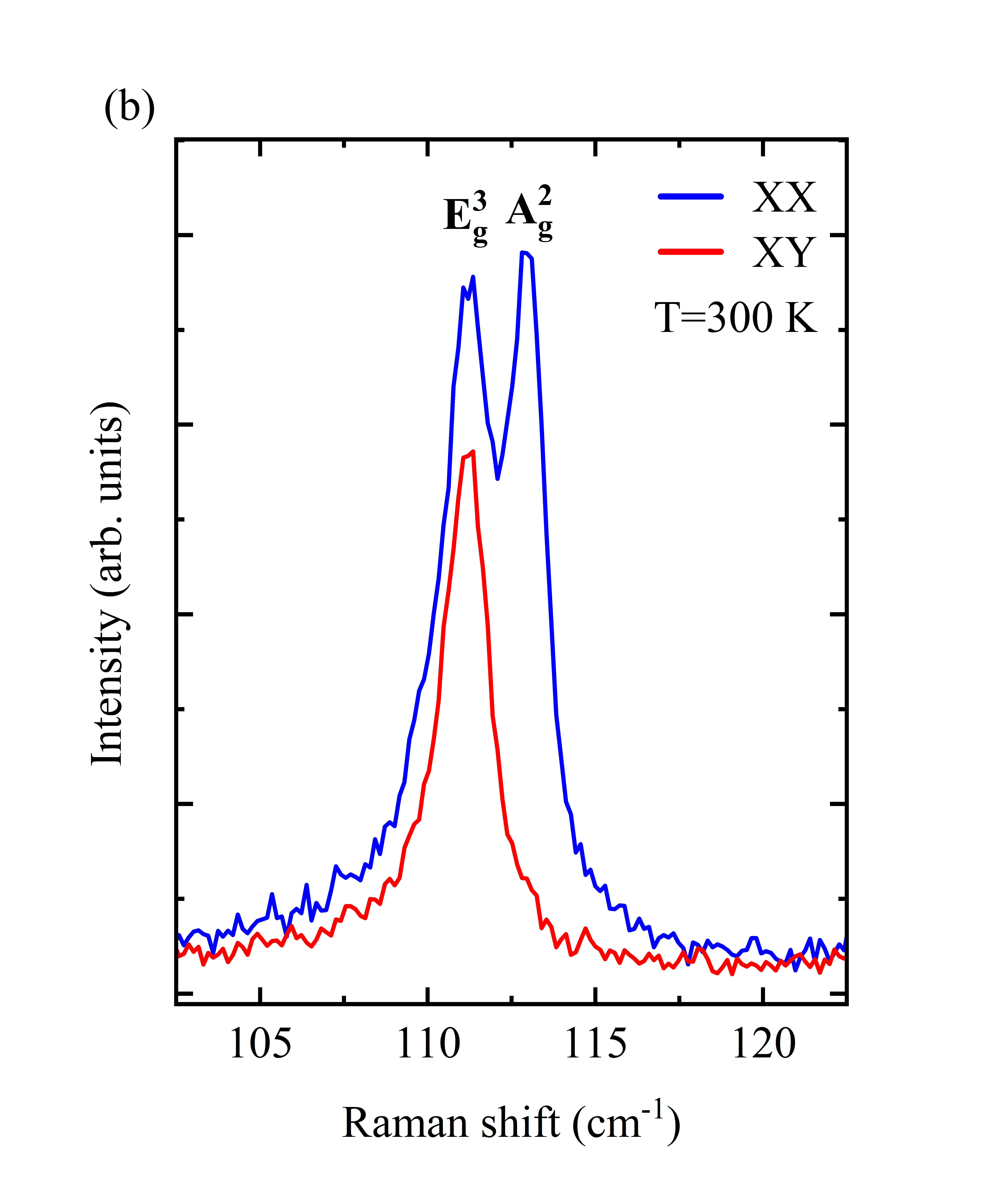}
    }
    \caption{(a) False-color map of the polarization sensitive RS spectra measured on the exfoliated 24-nm thick CGT flake at 5~K under excitation energy 1.58~eV and excitation power of 1~mW.  
    The white horizontal dot lines denote the co-linear (XX) and cross-linear (XY) polarizations. 
    (b) The corresponding RS spectra of the same CGT flake measured at 300~K in the energy range of the $\textrm{E}^3_\textrm{g}$ and $\textrm{A}^2_\textrm{g}$ peaks with XX and XY polarizations.}
	\label{map_polar}
\end{figure*}

Figure~\ref{map_polar}(a) shows the false-color map of the polarization-sensitive RS spectra measured on the 24-nm thick CGT flake.
Two types of Raman peaks are seen in the Figures.
The peaks ascribed to the $\textrm{A}_\textrm{g}$ modes are characterized by a strong dependence on the relative alignment of the excitation and detection polarization, leading to their appearance only for the co-linear configuration (XX).
For the $\textrm{E}_\textrm{g}$ peaks, they are visible in both co- (XX) and cross-linear (XY) configurations.

To unveil the polarization dependences in the vicinity of $\textrm{E}^3_\textrm{g}$ and $\textrm{A}^2_\textrm{g}$ peaks, exhibiting nearly degenerate energies at 5~K, we measured the polarization-resolved RS spectra of these modes at room temperature, see Fig.~\ref{map_polar}(b).
The energy overlap of the $\textrm{E}^3_\textrm{g}$ and $\textrm{A}^2_\textrm{g}$ modes is much smaller at 300~K , and hence two well-resolved lines are distinguishable, even in the XX configuration. 
Comparison of RS spectra in the XX and XY orientations results in a straightforward assignment of the apparent peaks to the $\textrm{E}^3_\textrm{g}$ and $\textrm{A}^2_\textrm{g}$ modes.

\newpage
\subsection{Analysis of the temperature evolutions the Raman peaks}

\begin{figure*}[!th]
		\subfloat{}%
		\centering
	\includegraphics[width=0.45\linewidth]{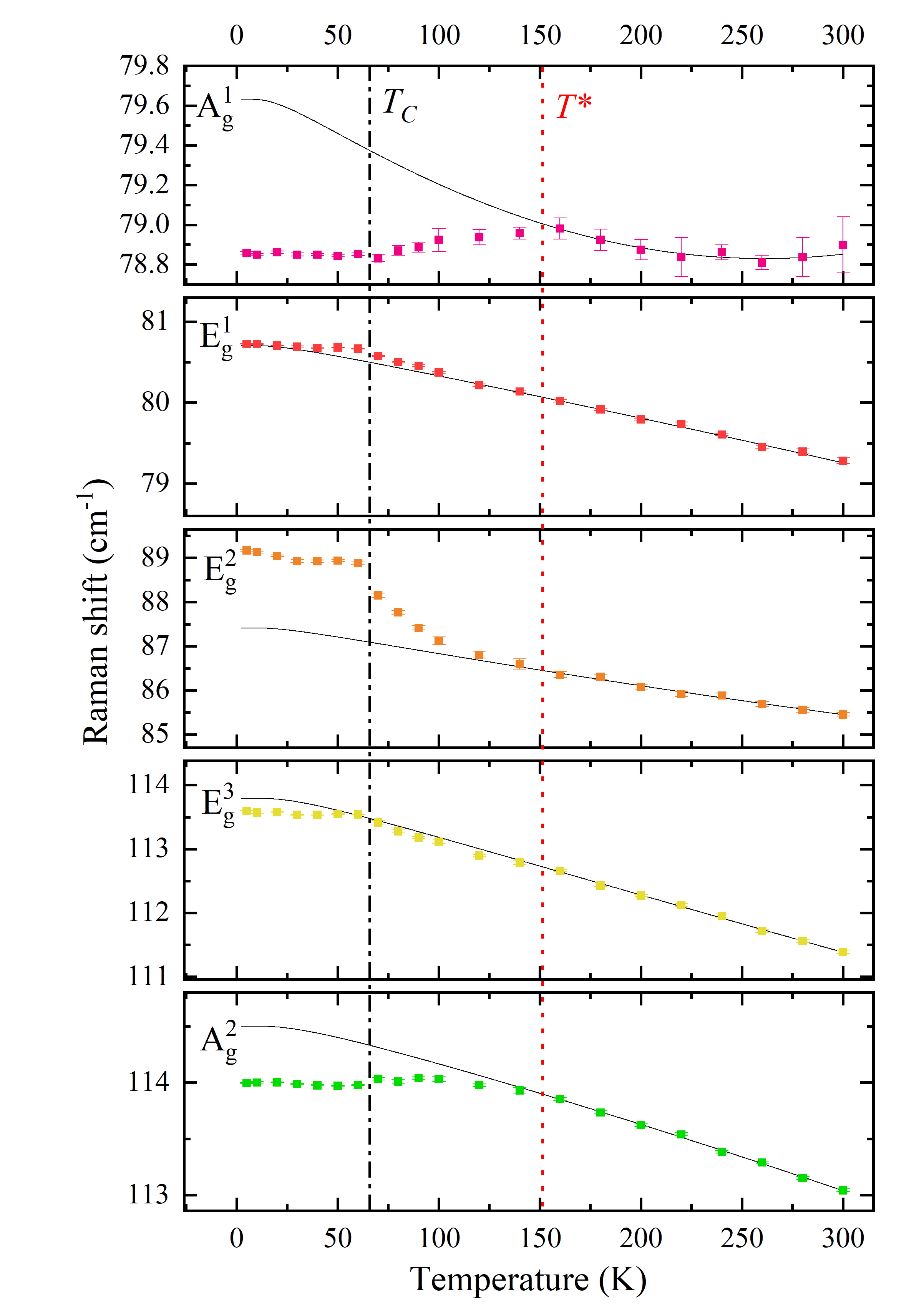}
        \includegraphics[width=0.45 \linewidth]{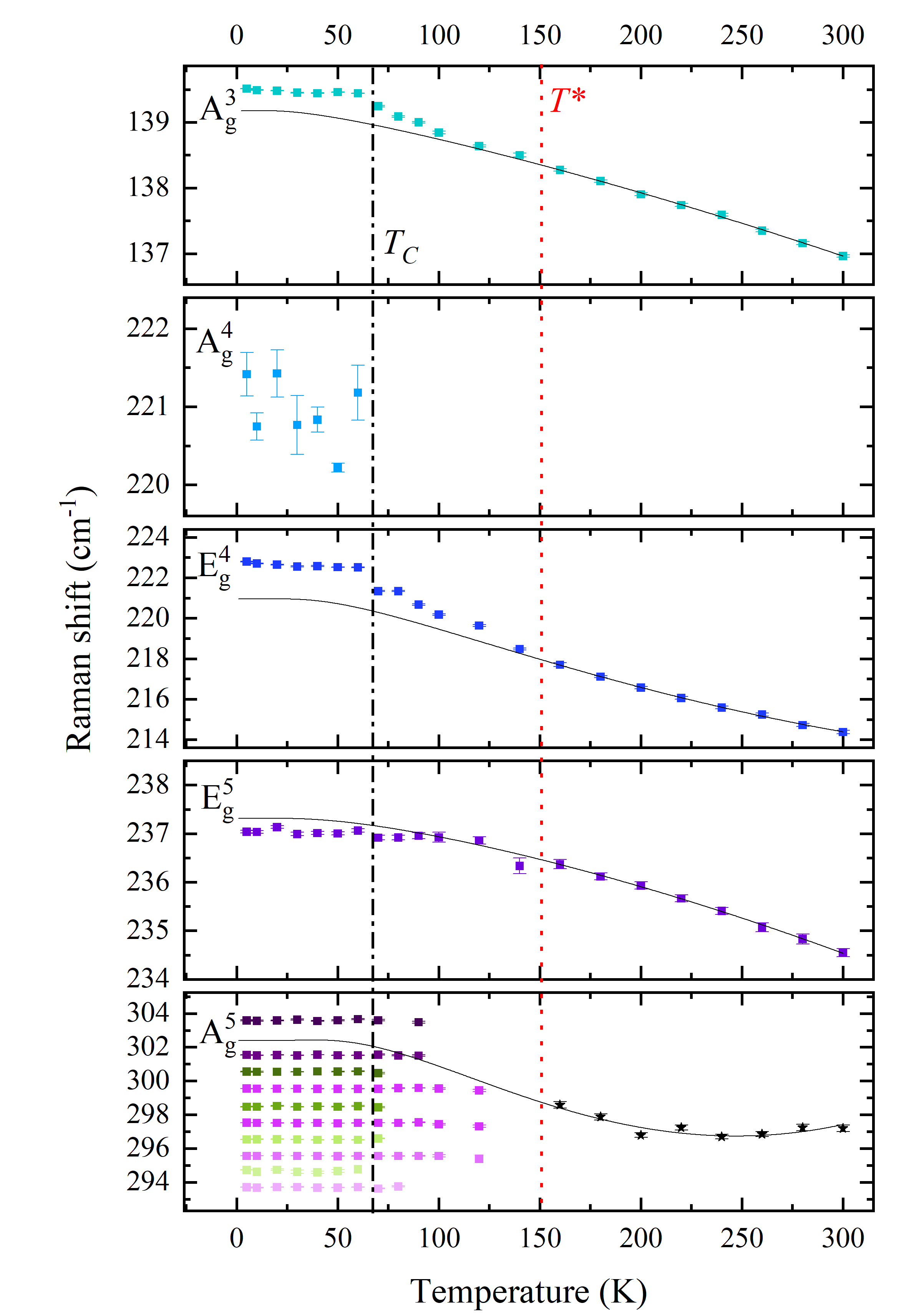}
        \caption{Temperature evolutions of the Raman shifts of the all investigated Raman modes measured on the exfoliated 24-nm thick CGT flake.
        The black vertical dash-dot line corresponds to the Curie temperature ($T_C$), while the red vertical doted line denote the second phase change temperature ($T^*$). 
        The gray curves are result of fitting using Balkanski model (Eg.~1 in the main text) to the data measured only above the $T^*$, $i.e.$ in the paramagnetic phase.}
		\label{E_phonon}
\end{figure*}

In the temperature evolution of the RS energies, several distinct behaviors can be identified. Within the ferromagnetic phase, the mode energies remain highly stable, exhibiting minimal temperature dependence. 
The A$_g^5$ mode in this regime displays its characteristic fine peak structure. 
Above the Curie temperature $T_C$, multiple types of behavior emerge. The A$_g^4$ mode cannot be identified anymore. 
The E$_g^2$, A$_g^3$, and E$_g^4$ modes exhibit a rapid, step-like redshift in energy, indicative of a strong positive SPC constant. The A$_g^1$ mode unexpectedly shows a slight blueshift to $T^*$. The fine structure of the A$_g^5$ mode gradually diminishes above $T_C$, with the intermediate array of peaks (colored green in Fig.~\ref{E_phonon}) becoming unresolvable.

The E$_g^1$, E$_g^3$, A$_g^2$, and E$_g^5$ modes show only a light change immediately after the phase transition, followed by a gradual redshift, consistent with expectations from phonon–phonon anharmonic scattering. 
At temperatures above $T^*$, the material enters a purely paramagnetic phase, and the energies of all Raman modes exhibit an anharmonic redshift. 
In this regime, the A$_g^5$ mode’s fine structure collapses into a single broad band.

We fitted the high-temperature data in the paramagnetic phase using the Balkanski model (presented in the main text), as indicated by the gray curves in Fig.~\ref{E_phonon}. Interestingly, the E$_g^3$, A$_g^2$, and E$_g^5$ modes, although exhibiting an apparently normal redshift, display a smaller energy shift than predicted by anharmonic scattering alone. 
This reduction in redshift may be a consequence of a negative SPC constant. 
Overall, the temperature dependence of the Raman mode energies reveals clear signatures of phase transitions driven by spin–phonon coupling.

\begin{table}[!th]
    \centering
    \begin{tabular}{|c|c|c|c|}
    \hline
Phonon mod	&	$\lambda$ (cm$^{-1}$)  &	$\lambda_1$(cm$^{-1}$) from Ref.~\citenum{Tian2016} &	$\lambda_2$(cm$^{-1}$) from Ref.~\citenum{Chakkar2024} 	\\
\hline
$\textrm{A}_\textrm{g}^1$   &	-0.34283	&		&		\\
$\textrm{E}_\textrm{g}^1$	&	0.01018	&		&	-0.4	\\
$\textrm{E}_\textrm{g}^2$	&	0.78144	&		&	-0.33	\\
$\textrm{A}_\textrm{g}^2$	&	-0.22347	&		&	-0.52	\\
$\textrm{E}_\textrm{g}^3$ &	-0.08567	&	0.24 &		\\
$\textrm{A}_\textrm{g}^3$	&	0.14873	&	0.32	&	-0.48	\\
$\textrm{E}_\textrm{g}^4$	&	0.82129	&	1.20&	-0.75	\\
$\textrm{E}_\textrm{g}^5$	&	-0.12279	&		&		\\

\hline

    \end{tabular}
    \caption{Spin-phonon coupling parameter calculated from fitting parameters of curves seen in \ref{E_phonon}. In latter columns are spin-phonon coupling factors found in literature, $\lambda_1$ from Ref.~\citenum{Tian2016} and $\lambda_2$ from Ref.~\citenum{Chakkar2024}. }
    \label{tab_spinphonon}
\end{table}

We calculated the spin-phonon coupling constants on the basis of Eq. 2 in the main text.
The extracted values are summarized in Table \ref{tab_spinphonon}. The coupling factors were successfully extracted for almost all modes, with two exceptions. 
For the $\textrm{A}_\textrm{g}^4$ mode, which disappears after the $T_C$ transition, it was not possible to determine the coupling constant. 
Similarly, for the $\textrm{A}_\textrm{g}^5$ mode in the ferromagnetic phase, the presence of a fine energy structure prevented the unambiguous selection of a single representative energy level for the calculation.

For the remaining modes, our results are largely consistent with those reported by Tian et al.~\cite{Tian2016} for $\textrm{E}_\textrm{g}^4$ and $\textrm{A}_\textrm{g}^3$. 
The $\textrm{E}_\textrm{g}^3$ mode exhibits a more pronounced deviation.
Due to the reported $\lambda$ values in Ref.~\cite{Chakkar2024} have large uncertainties, it makes their comparison with our extracted values impractical.

\begin{figure*}[!th]
		\subfloat{}%
		\centering
		\includegraphics[width=0.45 \linewidth]{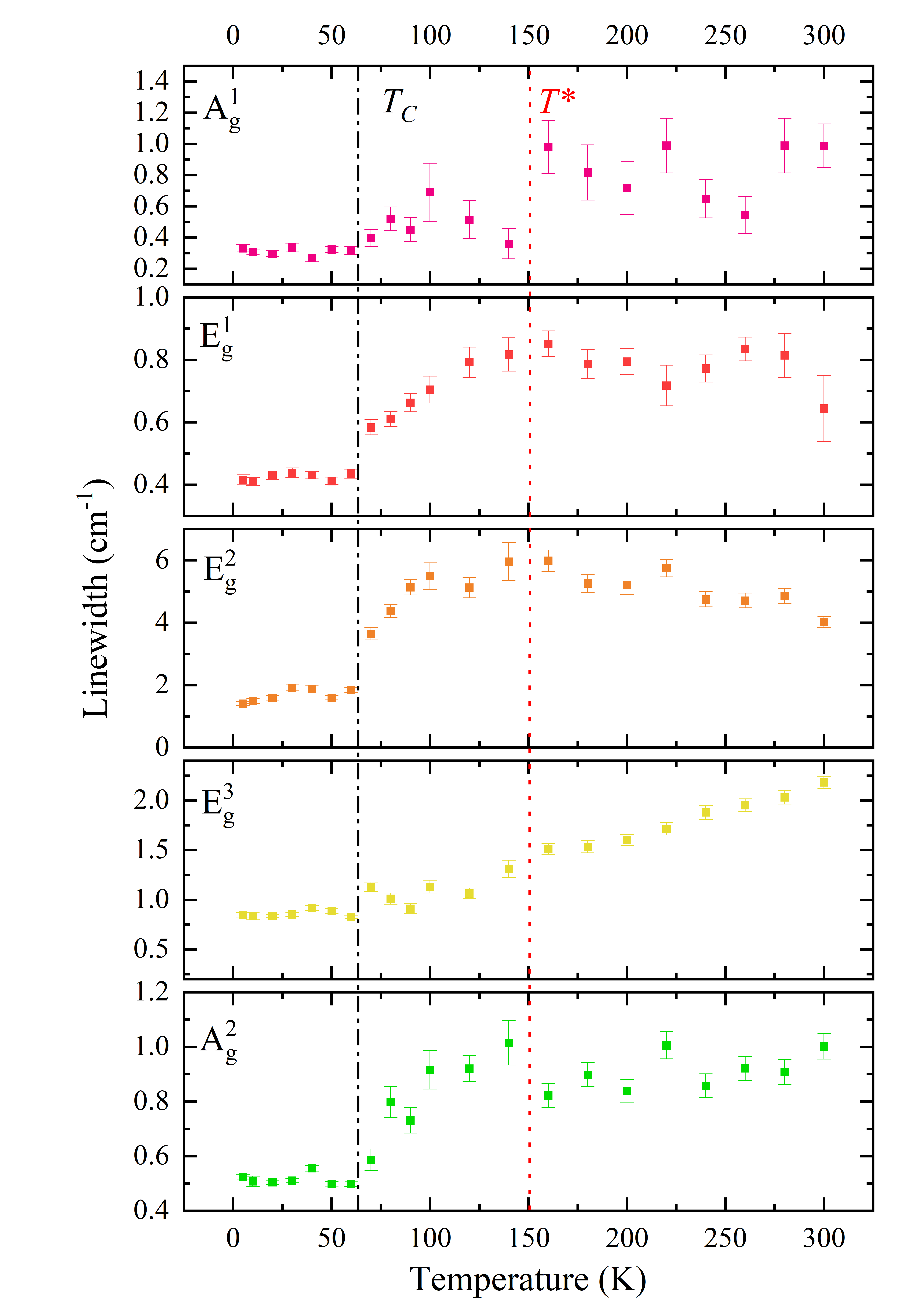}
   		\includegraphics[width=0.45 \linewidth]{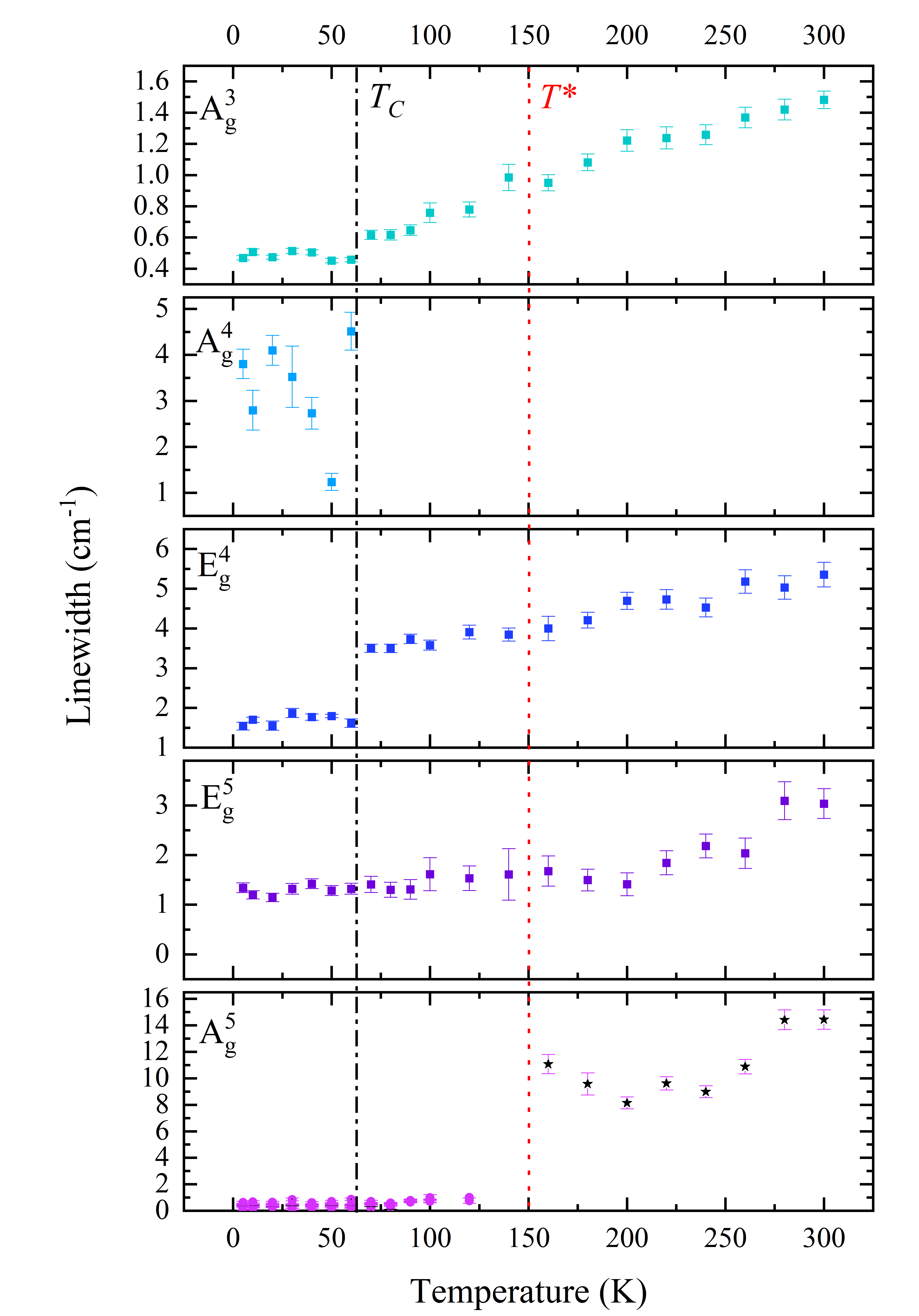}
        \caption{Temperature evolutions of the linewidths of the all investigated Raman modes measured on the exfoliated 24-nm thick CGT flake.
        The black vertical dash-dot line corresponds to the Curie temperature ($T_C$), while the red vertical doted line denote the second phase change temperature ($T^*$).}
		\label{ffwhm_phonon}
\end{figure*}

The temperature dependence of the linewidths, presented in Fig.~\ref{ffwhm_phonon}, further supports these observations. 
For almost all modes, a pronounced broadening occurs at the phase transition. 
In particular, the E$_\textrm{g}^1$, E$_\textrm{g}^2$, E$_\textrm{g}^3$, A$_\textrm{g}^2$, A$_\textrm{g}^3$, and E$_\textrm{g}^4$ modes show a clear step-like increase in linewidth at $T_C$. 
In particular, the E$_\textrm{g}^1$ mode, one of the few modes without a detectable SPC effect in its energy evolution, displays a strong SPC-related response in its linewidth behavior. 
Due to its disappearance above $T_C$, the A$_\textrm{g}^4$ mode could not be analyzed in this respect. 
The A$_\textrm{g}^1$ mode also shows an additional increase in linewidth above $T^*$, similar to the A$_\textrm{g}^5$ mode. 
This suggests that these modes are particularly sensitive to residual magnetic domains that persist above $T_C$.

\begin{figure*}[!th]
		\subfloat{}%
		\centering
		\includegraphics[width=0.45 \linewidth]{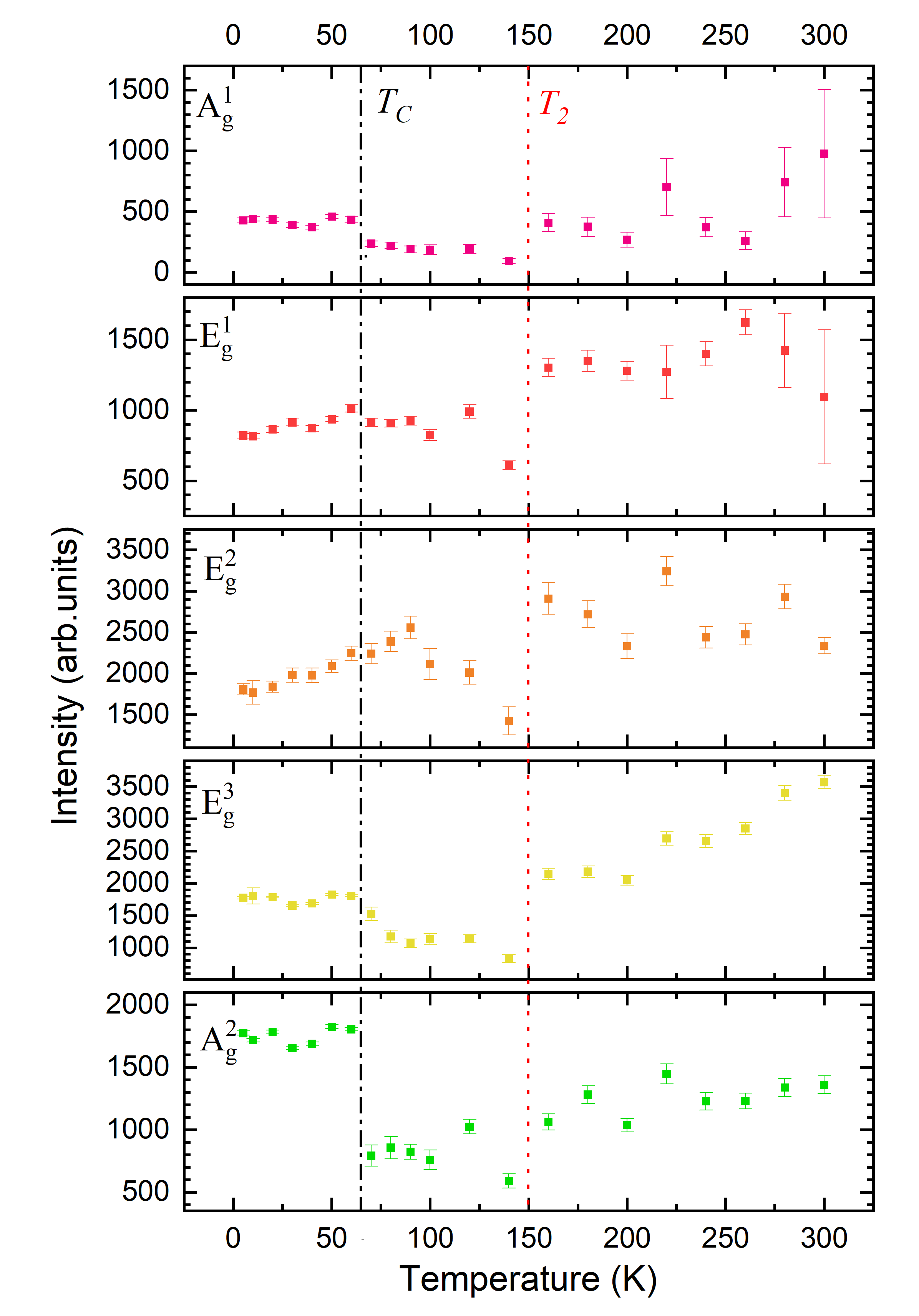}
        		\includegraphics[width=0.45 \linewidth]{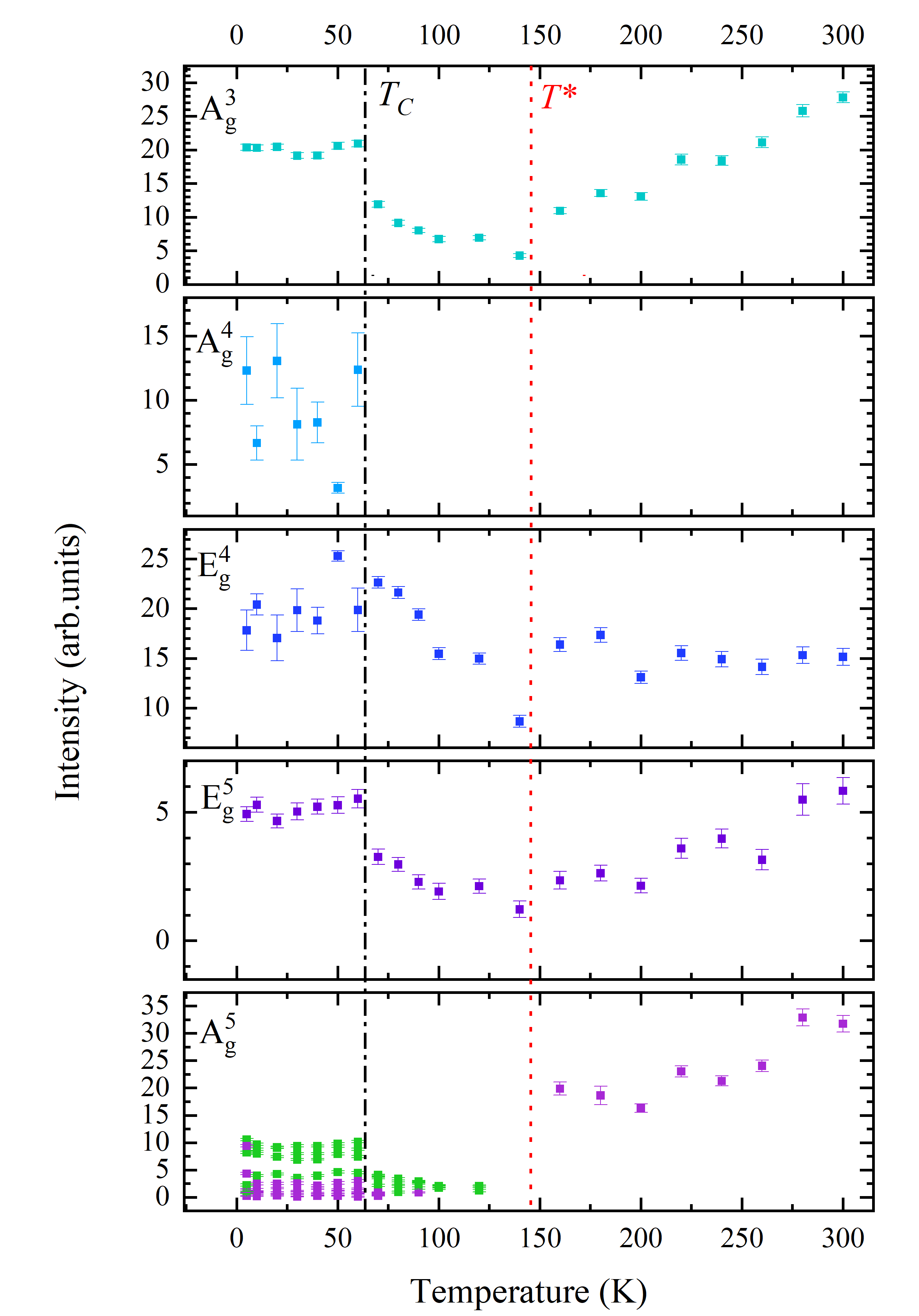}
        \caption{Temperature evolutions of the intensities of the all investigated Raman modes measured on the exfoliated 26-nm thick CGT flake. 
        The black vertical dash-dot line corresponds to the Curie temperature ($T_C$), while the red vertical doted line denote the second phase change temperature ($T^*$). }
		\label{int_phonon}
\end{figure*}

As seen in Fig.~\ref{int_phonon}, within the ferromagnetic phase, the intensities of most Raman modes remain relatively stable, exhibiting no significant variations. 
Upon entering the intermediate phase, just above the Curie temperature $T_C$, many modes experience a rapid decrease in intensity. This reduction is particularly pronounced for the A$_\textrm{g}^1$, A$_\textrm{g}^2$, A$_\textrm{g}^3$, E$_\textrm{g}^5$, A$_\textrm{g}^5$, and A$_\textrm{g}^4$ modes, with the latter disappearing entirely. 
This trend indicates that the intensities of the A$_\textrm{g}$ modes are generally more sensitive to the $T_C$ transition. 
In contrast, the E$_\textrm{g}^1$ and E$_\textrm{g}^3$ modes display a more gradual decrease in intensity.

Within the intermediate phase, the A$_\textrm{g}^1$, E$_\textrm{g}^1$, E$_\textrm{g}^3$, and A$_\textrm{g}^2$ modes maintain a nearly constant intensity, whereas the E$_\textrm{g}^2$, A$_\textrm{g}^3$, E$_\textrm{g}^4$, E$_\textrm{g}^5$, and A$_\textrm{g}^5$ modes show a steady decline up to $T^*$. 
Upon entering the purely paramagnetic phase above $T^C$, the intensities of all modes increase sharply, highlighting the influence of residual magnetic domains in the intermediate phase on the Raman scattering intensity.
\newpage
\subsection{Atomic displacements of A$_\textrm{g}^5$}
To validate the approximation of the A$_\textrm{g}^5$ Raman mode as an isolated Ge–Ge pseudo-molecular vibration, the $\Gamma$-point phonon eigenvector was extracted from density functional theory (DFT) calculations, and the corresponding atomic displacement amplitudes were quantified. As summarized in Table~\ref{SItab_displacement}, the vibrational dynamics are overwhelmingly dominated by the central Ge–Ge pair. The normalized displacement amplitude for each Ge atom is $\sim 0.697$, whereas the surrounding Te and Cr atoms exhibit substantially smaller displacements of $\sim 0.068$ and $\sim 0.005$, respectively. 
The Ge atomic displacements are thus roughly an order of magnitude larger than those of Te and more than two orders of magnitude larger than those of Cr. Furthermore, the Cartesian components of the calculated eigenvectors reveal that the two Ge atoms oscillate in anti-phase (moving in equal and opposite directions along the Ge–Ge bond axis), directly confirming a pure Ge–Ge stretching character. The negligible involvement of the surrounding Cr and Te sublattices confirms that host lattice interactions contribute minimally to this mode, providing direct microscopic justification for treating the A$_\textrm{g}^5$ vibration within an effective diatomic Ge–Ge pseudo-molecular framework.

\begin{table}[!th]
    \centering
    \begin{tabular}{|c|c|c|c|c|}
    \hline
	&	x	&	y	&	z	&	magnitude	\\
    \hline
Cr	&	0.004372	&	0.002224	&	0.001507	&	0.00513096	\\
Cr	&	-0.00437	&	-0.00222	&	-0.00151	&	0.00513096	\\
Ge	&	-0.59407	&	-0.30217	&	-0.20476	&	0.69725063	\\
Ge	&	0.594073	&	0.302174	&	0.204761	&	0.69725063	\\
Te	&	-0.04603	&	0.04944	&	-0.00649	&	0.06786082	\\
Te	&	0.046029	&	-0.04944	&	0.006486	&	0.06786076	\\
Te	&	-0.01717	&	-0.04444	&	0.048323	&	0.06786069	\\
Te	&	0.017166	&	0.044444	&	-0.04832	&	0.06786071	\\
Te	&	0.012851	&	-0.0306	&	-0.05919	&	0.06786065	\\
Te	&	-0.01285	&	0.030604	&	0.059189	&	0.06786069	\\
\hline
    \end{tabular}
    \caption{Cartesian coordinates and magnitude of atomic displacements in A$_\textrm{g}^5$ mode.}
    \label{SItab_displacement}
\end{table}

\newpage
\subsection{Thickness evolution of Raman spectra}

\begin{figure}[!th]
		\subfloat{}%
		\centering
		\includegraphics[width=0.8 \linewidth]{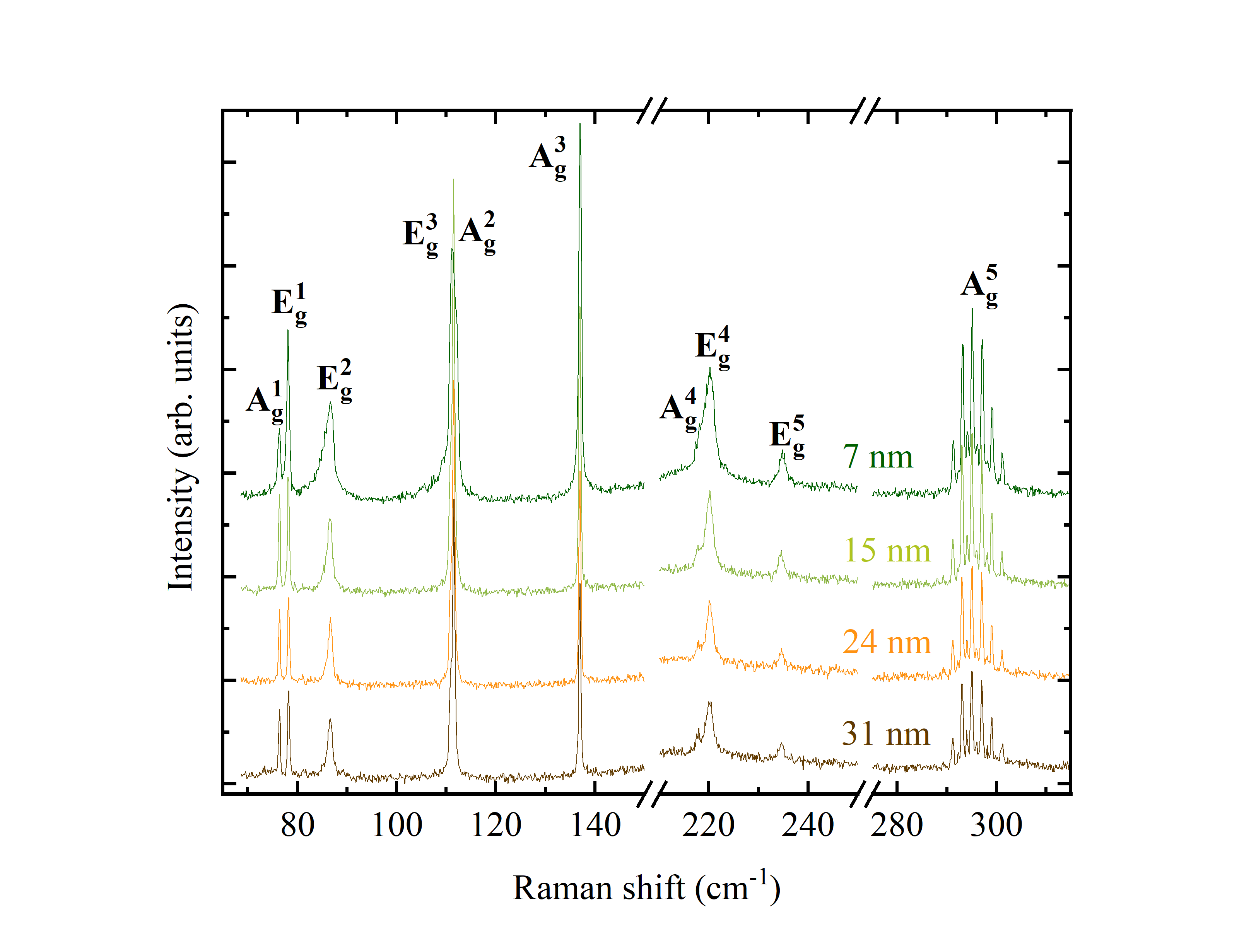}
    	\caption{RS spectra of the exfoliated thick CGT flakes with thicknesses: 7~nm, 15~nm, 24~nm, and 31~nm measured at 5~K with excitation energy 1.58~eV and excitation power 1~mW. 
        The spectra are vertically shifted for clarity.}
		\label{thickness}
\end{figure}

Figure \ref{thickness} presents Raman spectra from four flakes of different thicknesses, as measured by AFM in Section S1 of the SI. The thinnest flake, with a thickness of 7 nm, exhibits a higher overall spectral intensity compared to the 31 nm flake. 
This is precisely why the thinner flake was selected for the analysis of the fine structure of the A$_\textrm{g}^5$ mode, since the intensity of the intermediate peaks within that structure was also enhanced.

\newpage
\subsection{Theoretical simulation of the A$_\textrm{g}^5$ fine structure}

\begin{figure}[!th]
		\subfloat{}%
		\centering
		\includegraphics[width=1 \linewidth]{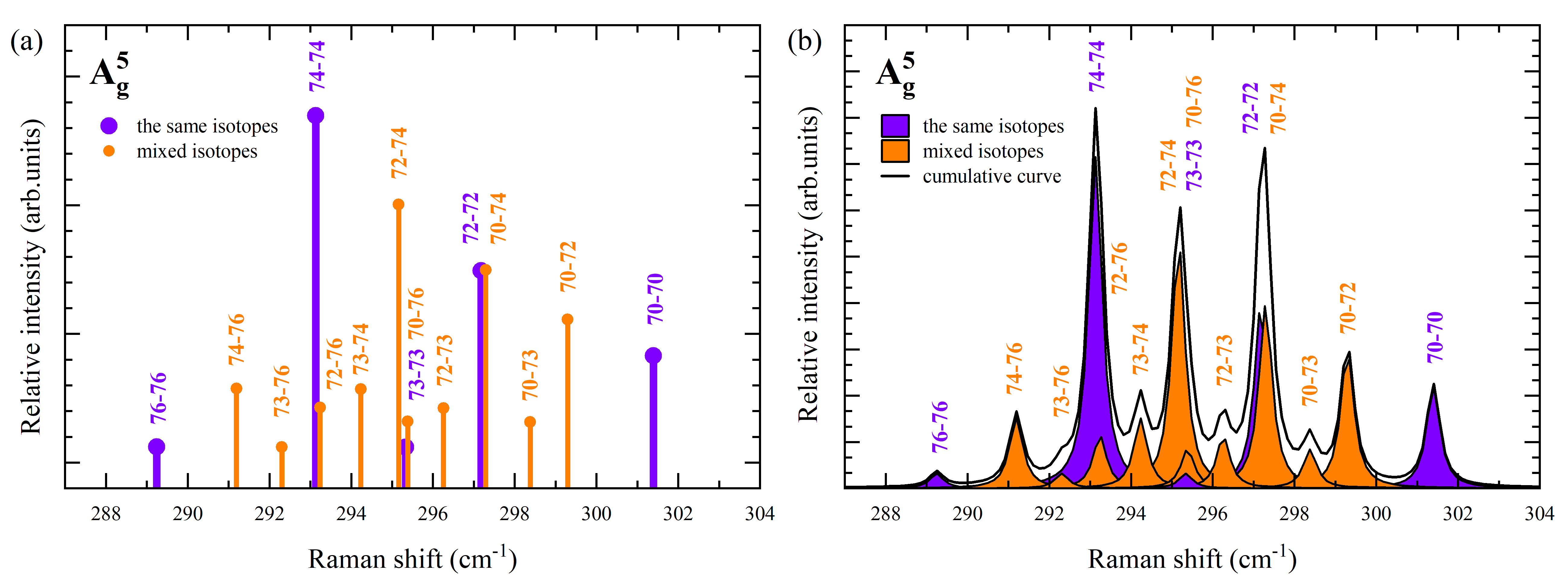}
    	\caption{Theoretical simulation of the RS spectrum of the A$_\textrm{g}^5$ mode including the isotopes related band structure: (a) without and (b) with additional broadening of the phonon modes. }
		\label{model}
\end{figure}

As described in the main text, the observed fine structure of the A$_\textrm{g}^5$ mode can be explained in terms of the Ge isotopes.
Therefore, we propose the following model. 
Due to the pseudo-molecule vibration of the Ge-Ge pair, we used the simple approach with two single-point masses connected by a spring. 
In this framework, the energy of the given vibration (in cm$^{-1}$) is given by
\begin{equation}
    E=\sqrt{\frac{K}{4\mu \pi^2 c^2}},
\end{equation}
where $K$ is the force constant, $\mu$ is the mass, and $c$ is the light velocity.
Note that such a model is typically applied to describe the so-called shear and breathing interlayer vibrations in thin layers on van der Waals materials, $e.g.$ transition metal dichalcogenides~\cite{Froehlicher2015, Grzeszczyk2016, Kipczak2020}.
In our simulations, we have made subsequent steps.
There are five isotopes apparent in natural Ge, 70, 72, 73, 74, and 76, with their corresponding abundances, 20.38$\%$, 27.31$\%$, 7.76$\%$, 36.72$\%$, and 7.83$\%$~\cite{deLaeter2003}.
Two kinds of Ge-Ge pairs are possible: five modes with the same isotope (70-70, 72-72, 73-73, 74-74, and 76-76) and ten modes with two different isotopes (70-72, 70-73, 70-74, 70-76, 72-73, 72-74, 72-76, 73-74, 73-76, and 74-76).
Taking into account that $E$ is inversely proportional to $\sqrt{\mu}$, we calculate the relative energy for different combinations of two Ge-Ge atoms with $\mu$ assumed as a reduced mass of two atoms (labeled $a$ and $b$) and reads $\mu=m_a*m_b/(m_a+m_b)$.
The relative intensity of each vibration was taken as probability of occurrence for peculiar pairs (the abundances squared for the same isotope, the product of abundance times two for dissimilar isotopes.
As the calculated energy of the spectrum is relative, we shifted its energy in a way that the energy of the 74-74 vibration matches the experimental energy of peak 1 from the A$_\textrm{g}^5$ mode (see Fig. 6 in the main text).
The determined spectrum of the A$_\textrm{g}^5$ fine structure composed of 15 peaks is shown in Fig.~\ref{model}(a).
In the final step, we simulate the Raman spectrum of the A$_\textrm{g}^5$ mode assuming the Lorentz distribution of a given peak with input of its frequency and intensity from Fig.~\ref{model}(a) and the determined experimental linewidth of 0.4~cm$^{-1}$, which is presented in Fig.~\ref{model}(b).

\end{document}